\documentclass[superscriptaddress,twocolumn,longbibliography]{revtex4-1}

\usepackage{amsmath}
\usepackage{amssymb}
\usepackage{amsxtra,color}
\usepackage{latexsym}
\usepackage{graphicx}
\usepackage[utf8]{inputenc}
\usepackage{MnSymbol}	% use for sumint

\usepackage{hyperref}

\definecolor{MyDarkGreen}{rgb}{0,0.6,0}
\definecolor{MyDarkBlue}{rgb}{0,0,0.8}
\definecolor{MyDarkRed}{rgb}{0.6,0,0.3}

\hypersetup{breaklinks=true, colorlinks=true,plainpages=true, linktocpage=true, linkcolor=MyDarkBlue, citecolor=MyDarkGreen, urlcolor=MyDarkRed, pdfborder={0 0 0},%
pdfauthor={},%
pdfsubject={Research article},
pdftitle={}%
}

\newcommand{\ba}{\begin{align}}
\newcommand{\ea}{\end{align}}
\newcommand{\be}{\begin{equation}}
\newcommand{\ee}{\end{equation}}

\newcommand{\ket}[1]{| #1 \rangle}

\newcommand{\braket}[2]{\langle \: #1 \: | \: #2 \: \rangle}
\newcommand{\braOket}[3]{\langle \: #1 \: | \: #2 \:| \: #3 \: \rangle}
\newcommand{\braOketred}[3]{\langle \: #1 \: \| \: #2 \: \| \: #3 \: \rangle}

\makeatletter
\newcommand*{\rom}[1]{\expandafter\@slowromancap\romannumeral #1@}
\makeatother
\newcommand{\EX}{E_\mathrm{X}}

\newcommand{\EL}{E_\mathrm{L}}

\newcommand{\cj}{c^{(1)}_{j'}}
\newcommand{\ck}{c^{(2)}_{j}}
\newcommand{\wSO}{\omega_\mathrm{SO}}
\newcommand{\ESO}{\epsilon_\mathrm{SO}}
\newcommand{\TSO}{T_\mathrm{SO}}

\newcommand{\tL}{\tau_\mathrm{L}^{(GD)}}

\newcommand{\tX}{\tau_\mathrm{X}^{(GD)}}
\newcommand{\phiX}{\phi_\mathrm{X}}
\newcommand{\threej}[6]{  \begin{pmatrix} #1 & #2 & #3 \\ #4 & #5 & #6 \end{pmatrix}}
\newcommand{\Cone}{\mathbf{C}^1}
\newcommand{\rfat}{\mathbf{r}}

\begin{document}

\title{Tutorial: Pulse analysis by delayed absorption from a coherently excited atom}

\author{Jan~Marcus \surname{Dahlstr\"om}}
\email{marcus.dahlstrom@matfys.lth.se}
\affiliation{Department of Physics, Lund University, Box 118, SE-221 00 Lund, Sweden}

\author{Stefan \surname{Pabst}}
\affiliation{ITAMP, Harvard-Smithsonian Center for Astrophysics, 60 Garden Street, Cambridge, MA 02138, USA}
\affiliation{Stanford PULSE Institute, SLAC National Accelerator Laboratory, Menlo Park, California 94025, USA}

\author{Eva \surname{Lindroth}}
\affiliation{Department of Physics, Stockholm University, AlbaNova University Center, SE-106 91 Stockholm, Sweden}

%\author{Eva \surname{Lindroth}}
%\affiliation{Department of Physics, Stockholm University, AlbaNova University Center, SE-106 91 Stockholm}

%32.80.-t : Photoionization and excitation
%32.80.Ee : Rydberg states
%31.15.vj : Electron correlation calculations for atoms and ions: excited states
%42.65.Re : Ultrafast processes; optical pulse generation and pulse compression 
\pacs{32.80.-t,42.65.Re,31.15.vj,32.80.Ee}
%32.80.Rm, 32.80.Qk, 42.65.Ky
%\date{}

\begin{abstract}
In this tutorial we provide a short review of attosecond pulse characterization techniques and a pedagogical account of a recently proposed method called Pulse Analysis by Delayed Absorption (PANDA) [Pabst and Dahlstr\"om, {\it Phys. Rev. A}, {\bf 94}, 13411 (2016)]. We discuss possible implementations of PANDA in alkali atoms using either principal quantum number wave packets or spin-orbit wave packets. 
The main merit of the PANDA method is that it can be used as a pulse characterization method that is free from atomic latency effects, such as  scattering phase shifts and long-lived atomic resonances. Finally, we propose that combining the PANDA method with angle-resolved photoelectron detection should allow for experimental measurements of attosecond delays in photoionization from bound wave packets on the order of tens of attoseconds.
\end{abstract}

\maketitle 

% Introduction:
\section{Introduction}
\label{sec:intro}

We live in a revolutionary time when quantum control of microscopic processes in matter is possible. Our increasing ability to control atoms and molecules is driven by rapid advances of laser light sources and there are currently several frontiers in laser-matter applications. 
%On the one hand, lasers with extreme precision in the spectral domain, with the aid of frequency combs, have been developed to reach frequency resolutions that allow for measurements of possible changes in the fundamental physical constants over time \cite{Hansch2006}. High-spectral laser precision also has opened up for atomic state preparation schemes that have been used for non-destructive measurements on single quantum systems \cite{Haroche2013}. On the other hand,
Optical laser pulses can be made ultra-short by decreasing the pulse duration down to the femtosecond barrier ($1$\,fs = $10^{-15}$\,s), limited only by the fundamental period of the laser light, for time-domain studies of molecular motion, known as {\it femtochemistry} \cite{Zewail2000}. The combination of pulse stretching, amplification and re-compression of laser light in the time domain has opened up for enormous increase of peak intensities \cite{Strickland1985} that can be used to ionize matter and drive electrons to high velocities. The electrons can be driven back to collide with their respective ions for coherent frequency conversion from low-frequency laser light to high-frequency light, in a processes called high-order harmonic generation (HHG) \cite{LewensteinPRA1994}, to form coherent bursts of radiation reaching the soft X-ray range \cite{Chen2010}. Given the right conditions and experimental filtering HHG can be used to form either trains of attosecond pulses \cite{PaulScience2001} or isolated attosecond pulses \cite{HentschelNature2001}. The duration of the individual attosecond pulses range from tens to hundreds of attoseconds ($1$ as = $10^{-18}$ s) and they are, therefore, said to ``break'' the femtosecond barrier that is inherent to optical pulses. Attosecond pulses are the shortest coherent light flashes created by man and can now be routinely generated in many laser laboratories around the world for the study of physics at the attosecond time scale, known as {\it attophysics} \cite{Krausz2009}. The fact that the attosecond pulses are naturally phase locked to a fundamental coherent laser field make them ideal for studies of electron dynamics in atoms and molecules by time-resolved pump--probe spectroscopy. The optical laser field can serve as a control field to either trigger ionization by quantum tunneling or to perturb undergoing electron dynamics by stimulated electron transitions. Recent experimental development has extended this control field to span from infrared, through visible, to the ultraviolet frequency range \cite{Wirth2011}, but future coherent non-linear experiments with multiple attosecond pulses hold even greater promise for control of the electron dynamics. As attophysics is a natural continuation of femtochemistry \cite{Zewail2000}, it is not surprising that many techniques have now been transferred from the femtosecond to the attosecond time domain. Such adaptations include both pulse characterization techniques \cite{MairessePRA2005} and studies of coherence properties of matter by transient absorption techniques \cite{GouliemakisNature2010}. The high frequency and large bandwidth of  attosecond pulses also opens up for core-specific transient absorption spectroscopy \cite{WornerNature2010}. One important difference between femtosecond and attosecond experiments is, however, that the high-photon energies, inherent to attosecond pulses, lead to ionization of matter and to the generation of photoelectrons. Naturally, a ``hot topic'' in attosecond physics is currently the determination of attosecond delays in photoemission in atoms~\cite{SchultzeScience2010,KlunderPRL2011,Ossiander2016,Isinger2017}, molecules~\cite{Huppert2016}, and solid state targets \cite{Cavalieri2007,Neppl2015}. 

At present time, experiments on the time scale of few attoseconds have only been possible by performing {\it relative} measurements between different targets, such as the relative photoionization delay between the $2s$ and $2p$ orbitals in neon \cite{SchultzeScience2010,Isinger2017}. The plain reason for this is that there exists currently no way to characterize attosecond pulses, without making severe approximations concerning the light--matter interaction, therefore, the unknown exact attosecond pulse shape in the experiment must by canceled out in some way \cite{DahlstromJPB2011}. Recently, Pabst and Dahlstr\"om have proposed a new type of scheme that holds the promise of {\it absolute} characterization of the attosecond pulses \cite{PabstPRA2016}. In this tutorial we will briefly review the state-of-the-art in attosecond pulse metrology and then explain our novel ideas of pulse characterization with numerical results for alkali atoms. We hope that this tutorial will help to improve attosecond pulse metrology with the aim to increase the temporal precision of future pump--probe experiments in physics, chemistry and material science.

\section{Transition from femtosecond to attosecond pulse characterization}

%The first thing that you should know about pulse characterization beyond the femtosecond barrier is that one does not measure light pulses in the temporal domain, but instead one measures photoelectron distributions in the energy domain. 
In order to characterize ultra-short light pulses one must determine both the spectral {\it magnitude} and {\it phase} of the pulses in the energy domain because the pulses are simply too short to be measured directly in the time domain. While the spectral magnitude can be easily obtained by linear spectroscopy, the determination of the spectral phase of the pulses is a much more demanding problem, which requires non-linear interactions of some sort. 
In the following, we will refer to the pulse that we wish to characterize as the {\it test pulse} denoted with subscript $X$. The electric field of the test pulse in the time domain can be expressed in terms of its Fourier components, 
\begin{align}
\tilde \EX(t)=\frac{1}{2\pi}\int_{-\infty}^{\infty} d\omega \EX(\omega)\exp[-i\omega t],
\label{EXoft}
\end{align}
where the spectral magnitude and phase can be separated as $\EX(\omega)=|\EX(\omega)|\exp[i\phiX(\omega)]$. The integral over angular frequency runs over negative and positive frequencies with the relation $\EX^*(\omega)=\EX(-\omega)$ to ensure that the physical electric field is a real function in time.
The detailed temporal structure of the test pulse is most conveniently described by its group delay,  
\begin{align}
\tX(\omega)=
\frac{\partial \phiX}{\partial \omega}, 
\end{align}
which is the spectral derivative of the spectral phase. Physically, the group delay of a certain angular frequency corresponds to the time of arrival of that specific frequency component of the test pulse at a given target. Given the group delay of a test pulse the spectral phase can be reconstructed, for instance, by integration from the central frequency of the pulse,  
\begin{align}
\phiX(\omega)=
\int^{\omega}_{\omega_X} d\omega' 
\tX(\omega')+\phi_0,
\end{align}
up to a constant spectral phase term, $\phi_0$. Physically $\phi_0$ determines the carrier-envelope phase (CEP) of the pulse in the time domain. While CEP effects have proven to be important to understand tunnel ionization from atoms and other phenomena driven by intense ultra-short laser fields \cite{Krausz2009}, CEP effects are absent in one-photon ionization and will not be considered in this work on attosecond pulse characterization. %, which is focused on attosecond pulse characterization in the XUV regime, where non-linear effects are weak given present-day pulse parameters. 

In the optical domain phase measurements can be carried out by different techniques including Frequency Resolved Optical Gating (FROG) \cite{FROG},  Spectral Phase Interferometry for Direct E-field Reconstruction (SPIDER)  \cite{Iaconis1998} and phase retrieval from second-harmonic dispersion scans (the d-scan method) \cite{Miranda:12}. All approaches rely on {\it parametric} non-linear optical processes, which means that the quantum system is returned to the same initial state after interaction with the field, e.g. after absorption of two laser photons and emission of one second harmonic photon \cite{Boyd}. Any excitation by the laser field is {\it virtual} so that no remaining excitation is possible. In parametric processes, no additional phase is introduced by the measurement process and the optical pulses can be characterized {\it exactly} in principle. 

\begin{figure}[!htb]
\includegraphics[width=0.5\textwidth]{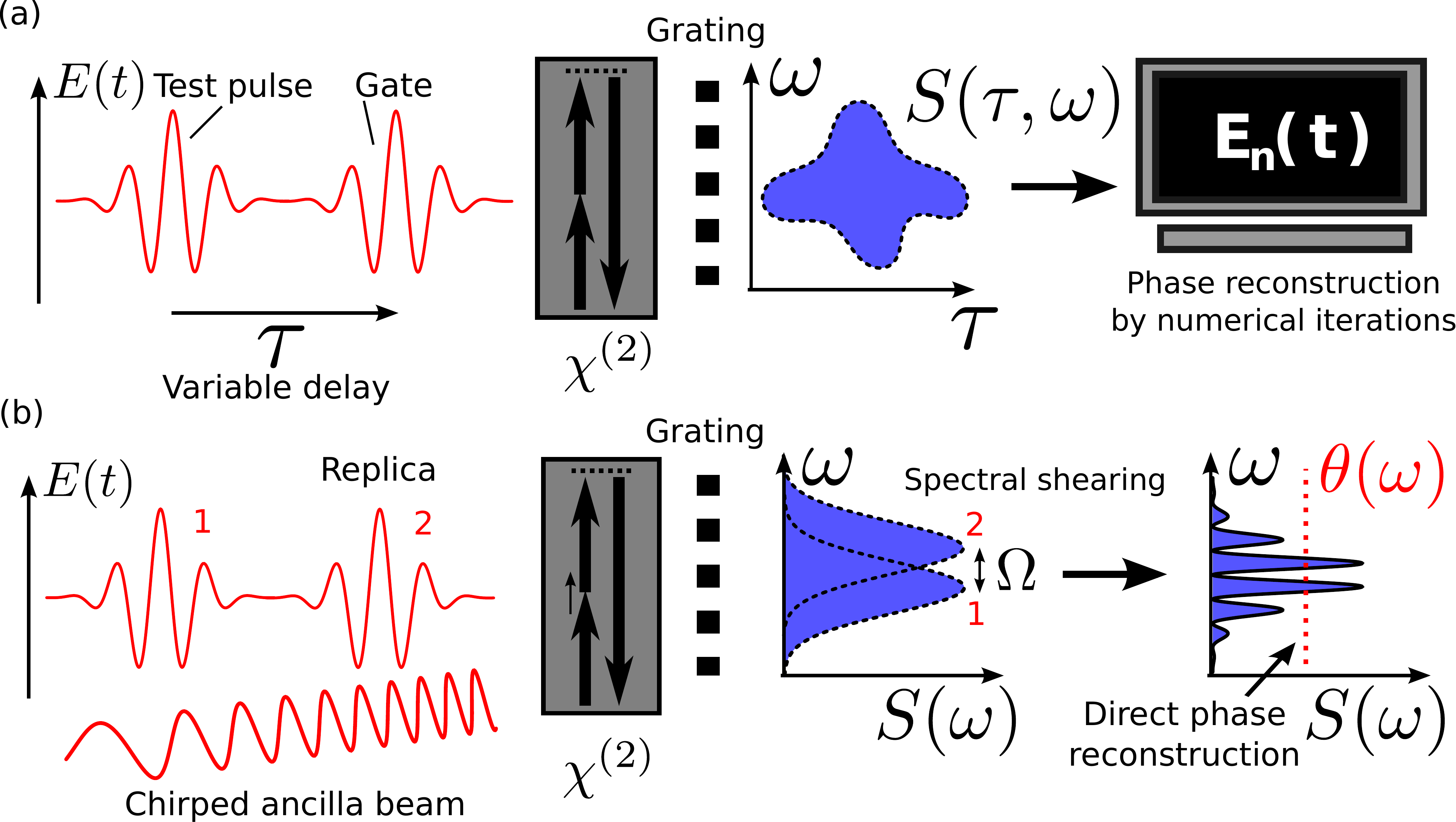}
\caption{
Schematic pictures of (a) the FROG method (b) the SPIDER method for pulse characterization of ultra-short laser pulses. 
\label{fig:frog-spider}}
\end{figure}

The principle for the FROG technique is shown in Fig.~\ref{fig:frog-spider}~(a), where a pulse replica serves as a ``gate'' to probe the structure of the optical test pulse. An auto-correlation spectrogram measurement of the second harmonic emission ($\chi^{(2)}$ process), resolved over both angular frequency, $\omega$, and time delay, $\tau$, provides sufficient information to reconstruct the pulse shape of the electric field in temporal domain by numerical iterations \cite{FROG}. 
The principle of the SPIDER technique is shown in Fig.~\ref{fig:frog-spider}~(b), where spectral-shearing interferometry (interference of different frequencies of the light pulse) is generated between two pulse replicas by non-linear mixing with a chirped ancilla beam ($\chi^{(2)}$ process) \cite{Iaconis1998}. The spectral phase (up to a constant) can be directly read out from a single SPIDER measurement without the need for numerical iterations and delay scans. 
This is in contrast to the FROG technique, where numerical iterations and delay scans are always required.  

\begin{figure}[!htb]
\includegraphics[width=0.5\textwidth]{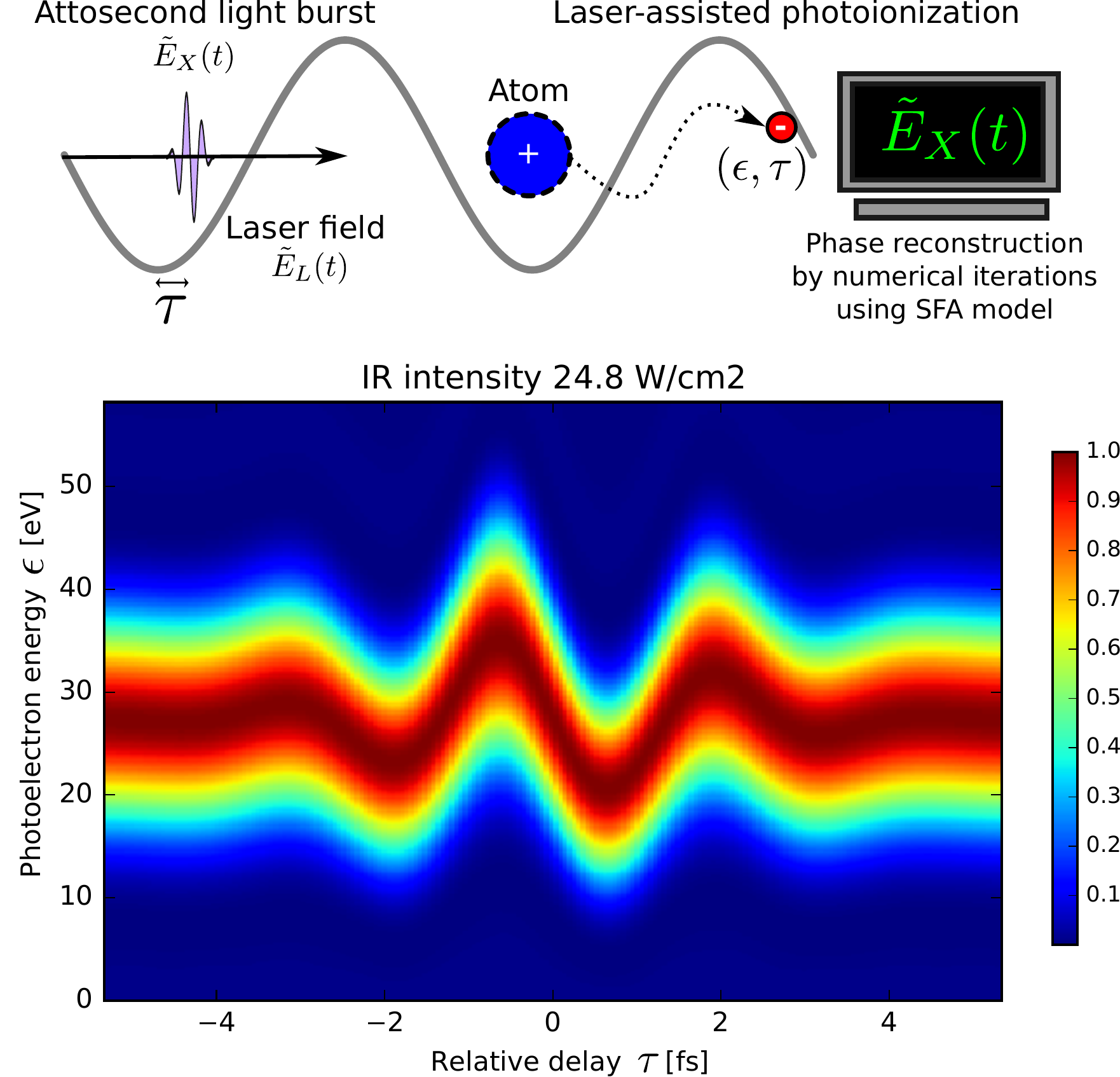}
\caption{
Schematic pictures of the FROG-CRAB method where an attosecond pulse is used to ionize an atom in the presence of a laser field (upper panel). The laser-assisted photoelectrons are recorded and used for pulse reconstruction by numerical iterations based on a simplified model of the complex ionization process. A generic streaking photoelectron spectrogram corresponding to a Fourier limited attosecond pulse is computed using the Strong Field Approximation (SFA) and shown in false colors in the lower panel. For simplicity the dipole transition matrix elements to the continuum are assumed to be one.  
\label{fig:streaking-simple}}
\end{figure}

Similar to the FROG technique, temporal information of attosecond pulses can be gained by non-linear cross-correlation with ultra-short laser pulses. The particular case of laser-assisted photoionization by an isolated attosecond pulse is called ``attosecond streaking'', because the final photoelectron momentum, $\mathbf{p}$,  is deflected by the laser field, $\mathbf{E}_\mathrm{L}(t)$, as determined by classical mechanics, 
\begin{align}
\mathbf{p} \approx \mathbf{p}_0 -e\int_{t_0}^{\infty} dt \mathbf{E}_\mathrm{L}(t) = \mathbf{p}_0-e\mathbf{A}_\mathrm{L}(t_0),
\label{poft}
\end{align}
where $\mathbf{p}_0$ is the initial momentum and $\mathbf{A}_\mathrm{L}(t_0)$ is the vector potential of the laser field at the ionization time triggered by the attosecond pulse \cite{ItataniPRL2002}. In this way, the femtosecond oscillations of an optical pulse, $A_\mathrm{L}(t)$, can be measured directly in the time domain, using a much shorter attosecond pulse that acts as a well-defined {\it amplitude gate} in the time domain \cite{GoulielmakisScience2004}. In contrast, Mairesse and Qu{\'e}re proposed to use the laser field as a {\it phase gate} to perform Frequency Resolved Optical Gating for Complete Reconstruction of Attosecond Bursts (FROG-CRAB) \cite{MairessePRA2005}, as illustrated in the upper panel of Fig.~\ref{fig:streaking-simple}. 
In the lower panel of Fig.~\ref{fig:streaking-simple}, we illustrate a  photoelectron spectrogram for a Fourier limited attosecond pulse, which is streaked by an ultra-short laser field, $A_\mathrm{L}(t)$. While the energy oscillations of the photoelectron in Fig.~\ref{fig:streaking-simple} are related to the laser pulse by Eq.~(\ref{poft}), the actual shape of the attosecond pulse is encoded in the spectrogram in a more complicated way and it is {\it not} directly observable by the naked eye.  
It is appealing to model laser-assisted photoionization with the strong field approximation (SFA), where the continuum states are approximated by Volkov states \cite{Kitzler2002}. Using SFA, the complex amplitudes for laser-assisted photoionization can be calculated easily by first-order time-dependent perturbation theory (here in velocity gauge) as 
\begin{align}
c_{\mathbf{k}i} &= \frac{e}{i m}
\braOket{\mathbf{k}}{k_z}{i}
\int_{-\infty}^{\infty} dt'
A_\mathrm{X}(t') 
\exp \left[\frac{i}{\hbar}S(t';\mathbf{k})\right]
\label{cofk}
\end{align}
where $A_\mathrm{X}$ is the vector potential of an attosecond pulse(s) with linear polarization along the $z$ direction and the action (or instantaneous energy) is 
\begin{align}
S(t';\mathbf{k})&=
\int^{t'} dt'' \left[
\frac{[\hbar\mathbf{k}+e\mathbf{A}_\mathrm{L}(t'')]^2}{2m}
+I_p \right],
\label{Sofk}
\end{align}
where $I_p$ is the binding energy of the atom and $\mathbf{A}_\mathrm{L}$ is the laser field used to streak the photoelectron. The one photon matrix element for photoionization from the $1s$ ground state of hydrogen to a plane-wave state is given by \cite{bethe-salpeter}
\begin{align}
\braOket{\mathbf{k}}{k_z}{i}=k_z\braket{\mathbf{k}}{i}=
\frac{2^{3/2}}{\pi}
\beta^{5/2}
\frac{k\cos\theta_k}
{(k^2+\beta^2)^2} \sim k^{-3}, 
%\\
%\braOket{\mathbf{k}}{z}{i}=
%i\frac{2^{7/2}}{\pi}\beta^{5/2} 
%\frac{k\cos\theta_k}
%{(k^2+\beta^2)^3}
\end{align}
with $\beta=Z/a_0$ where the nuclear charge $Z=1$ for hydrogen and $a_0$ is the Bohr radius. 
In this way Mairesse and Qu{\'e}re used the SFA to adopt the established FROG technique from ultra-fast laser optics \cite{FROG} to reconstruct simultaneously {\it both} the attosecond pulse(s) and the laser probe field \cite{MairessePRA2005}. 
This type of laser-assisted photoionization forms the basis for all attosecond pulse characterization at present time and other related techniques include Phase Retrieval by Omega Oscillation Filtering (PROOF) \cite{ChiniOE2010}, which is used for isolated attosecond pulses with the perturbative laser field, and RABBIT \cite{PaulScience2001}, which is used for characterization of periodic trains of attosecond pulses with a perturbative laser field.  %In order to perform FROG-CRAB, PROOF or RABBIT measurements one must  generate a complete photoelectron spectrogram over electron-energy and relative delay between the attosecond pulse and the laser pulse. 

%\subsection{The breakdown of the FROG-CRAB technique}
There is a general problem of quantitative accuracy of any laser-assisted photoionization technique (such as FROG-CRAB) when the SFA is used for the description of the photoionization process due to the omitted short-range and long-range interactions between electron and ion \cite{DahlstromJPB2012}. The first real proof of the ``breakdown'' of the FROG-CRAB technique for atomic targets was shown in a relative delay experiment between the neon orbitals $2p$ and $2s$ by Schultze {\it et al.} in 2010 \cite{SchultzeScience2010}. Despite numerous theoretical simulations, satisfactory agreement between theory and experiment has still not been reached, as reviewed by Feist {\it et al.} \cite{FeistPRA2014}. 
%This ``breakdown'' of FROG-CRAB was intriguing and inspired 
Complementary intra-atomic delay studies have been performed using the RABBIT technique between the $3p$ and $3s$ orbitals in argon atoms \cite{KlunderPRL2011,GuenotPRA2012}. Inter-atomic measurements between different noble gas atoms have also been reported \cite{PalatchiJPB2014,GuenotJPB2014} as well as relative measurements between single and in double photoionization \cite{ManssonNP2014}. 
Theoretically, the delays that appear in laser-assisted photoemission due to atomic interactions, $\tau_\mathrm{A}$, can be interpreted as the sum of two terms, 
\begin{align}
\tau_A\approx\tau_{w}+\tau_{cc},
\label{tauA}
\end{align}
where $\tau_w$ is the one-photon Wigner-like delay of the photoelectron \cite{WignerPR1955} and $\tau_{cc}$ is the continuum--continuum delay (also called Coulomb-Laser Coupling) that arises in the laser-stimulated transition between two continuum states in the presence of a long-range Coulomb potential \cite{KlunderPRL2011,DahlstromJPB2012,DahlstromCP2013,NagelePRA2011,PazourekPRL2012,PazourekFD2013,Lindroth2017}. It is possible to motivate why FROG-CRAB, PROOF and RABBIT experiments can be all interpreted by Eq.~(\ref{tauA}) using the asymptotic phase shifts of two-photon matrix elements \cite{DahlstromCP2013}, or  by direct numerical simulations \cite{PazourekFD2013}. This new interpretation of laser-assisted photoionization has been recently confirmed using isolated attosecond pulses, to extract the delay of helium relative to spectrally resolved shake-up satellites, by Ossiander {\it et al.} \cite{Ossiander2016}; and also by attosecond pulse trains, to extract the relative $2p-2s$ delay of neon, by Isinger {\it et al.} \cite{Isinger2017}. In the work of Isinger {\it et al.} we stress that the  combined high-spectral and temporal resolution provided by attosecond pulse trains was essential to spectrally remove nearby shake-up satellites. The influence of shake-up satellites can be expected to increase as the central frequency and bandwidth of extreme light pulses grow, thus, the problems associated with shake-up satellites is likely to increase the general uncertainty of the FROG-CRAB technique \cite{PabstJPB2017}. 
In hindsight, it is now clear that ``cumbersome'' laser-assisted photoionization simulations, including electron--electron correlation effects, are required to  quantitatively model attosecond streaking experiments in noble gas atoms. These type of simulations can only be performed with great precision in helium (the simplest noble gas atom) \cite{Ossiander2016} or by numerically efficient methods based on atomic many-body perturbation theory \cite{DahlstromPRA2012,Lindroth2017} or by large-scale numerical propagation of a suitably truncated many-body basis \cite{MoorePRA2011}. 

Laser-assisted photoionization is not the only method that has been proposed for characterization of attosecond pulses. Another method, which was used in early attempts to determine sub-femtosecond pulse structures, is based on  autocorrelation measurements \cite{Tzallas2003}. This type of non-linear measurements at short-wave lengths are very challenging experimentally and limited to rough pulse duration estimates \cite{Thomson2013}. 
Another proposal for attosecond pulse characterization is based on {\it in-situ} measurements to probe the ``birth'' of attosecond pulses, as proposed in 2006 by Dudovich {\it et al.} \cite{Dudovich2006}. The general idea is to gently perturb the HHG process and perform electron interferometry by inducing phase differences on the electron trajectories. This was first proposed to be done by a weak {\it parallel} second harmonic laser field, but it was later shown that this weak field strongly affected the initial tunneling step of HHG, and the subsequent interference pattern of the {\it in-situ} scheme \cite{DahlstromJPB2011}, which made the method unfit for accurate attosecond pulse reconstruction. In contrast, a more recent {\it in-situ} scheme based on two laser fields with {\it orthogonal} polarization has been used to study electron trajectories of the HHG process \cite{ShafirNature2012}. Despite this latter success on the microscopic scale, {\it in-situ} measurements remain of limited use for pulse characterization in general, because they can not be used to determine the final shape of the macroscopic pulse on a given target far away from the HHG site, which is affected by pulse propagation in the extended gas medium \cite{GaardeJPB2008} and general dispersion and absorption due to any optical elements in the beam path after HHG. 

For all these reasons we believe that a new scheme for attosecond pulse characterization, which is based on a simpler physical process, is in order and we will discuss our proposal for a solution in next section.

\section{The PANDA method}
In this section we discuss the use of a bound time-dependent wave packet as a ``clock'' for attosecond pulse characterization. We refer to this technique as Pulse Analysis by Delayed Absorption (PANDA) \cite{PabstPRA2016} and we stress that it is distinct from the above mentioned characterization techniques because it relies on {\it sequential} photoionization rather than laser-assisted photoionization. Here, we focus on the application of PANDA to attosecond pulses, but it is possible to use the scheme to characterize long-wavelength pulses, such as ultra-short optical laser pulses, and short-wavelength narrow-bandwidth pulses, such as Free Electron Laser (FEL) pulses, by suitable design of the bound wave packet, see Sec.~\ref{sec:designwp}.  

\begin{figure}[!htb]
\includegraphics[width=0.5\textwidth]{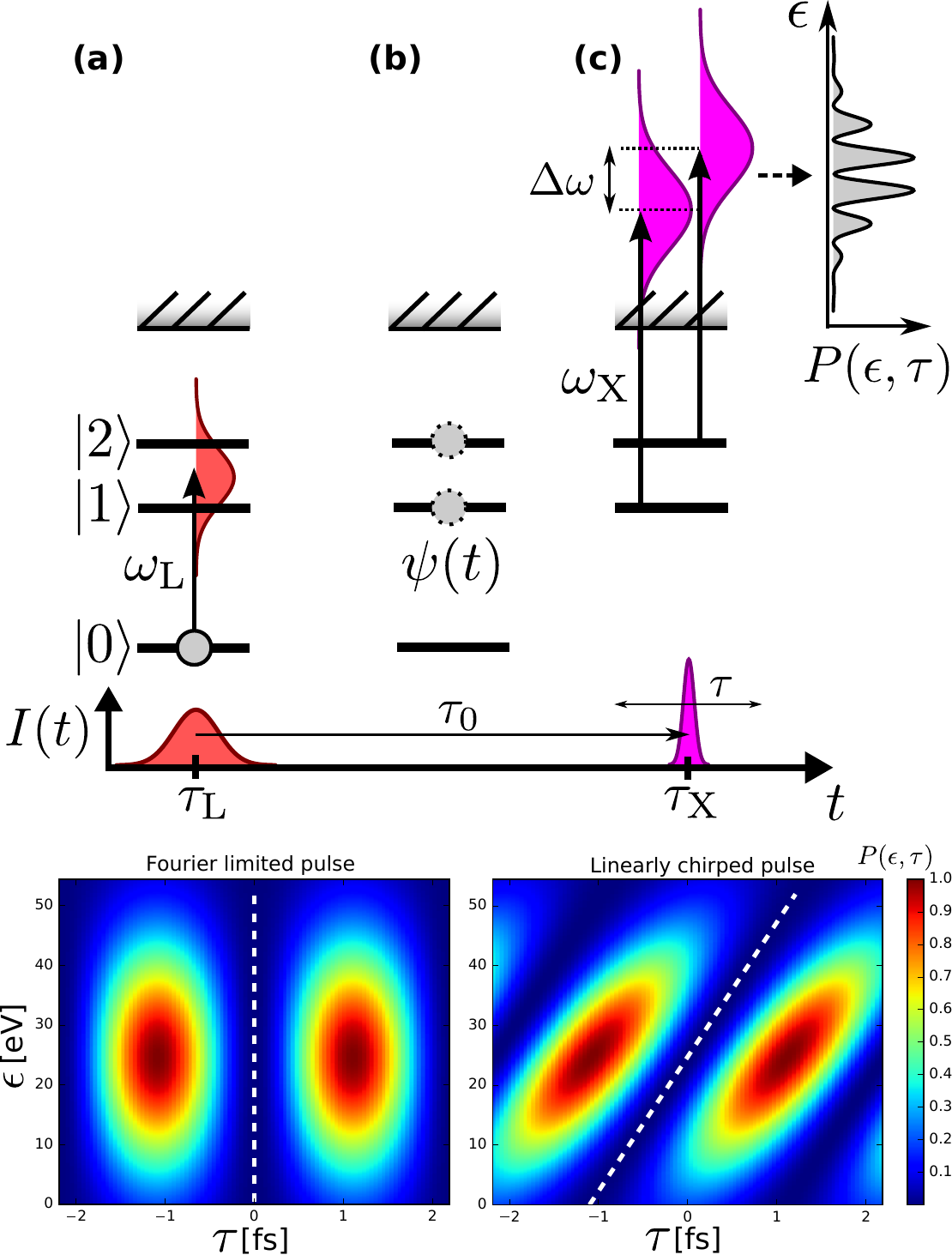}
\caption{
The three basic steps of the PANDA method for attosecond pulse characterization. (a) Coherent preparation of a bound wave packet that consists of states $\ket{1}$ and $\ket{2}$, (b) free evolution of the wave packet for a time $\tau$ and (c) spectral shearing interferometry of the attosecond pulse (X) by energetically overlapping photoionization processes from wave packet states with angular frequency splitting $\Delta\omega$. 
In the lower panels we show photoelectron spectrograms $P(\epsilon,\tau)$ for the PANDA method using a Fourier limited attosecond pulse (left) and a linearly chirped attosecond pulse (right). For simplicity the dipole transition matrix elements to the continuum are assumed to be one. The white dashed line shows the corresponding group delay of the attosecond pulse in excellent agreement with the quantum beating of the PANDA spectrogram (shifted to the node of the beating for clarity).   
\label{fig:basicidea}}
\end{figure}

The three basic steps of the PANDA method are illustrated in Fig.~\ref{fig:basicidea}. In step (a), a coherent bound wave packet is created, for instance by an intense laser pulse with central frequency, $\omega_\mathrm{L}$. In step (b), the bound wave packet, $\psi(t)$, is freely propagated for a controllable time, $\tau$. In step (c), the bound wave packet is ionized by one-photon absorption from the attosecond pulse with the ejected electron collected over all emission angles as a function of kinetic energy, $\epsilon$. Finally, the process is repeated for other (sequential) delays, $\tau$, to clearly resolve quantum beating in a photoelectron spectrogram, as illustrated in the lower panels of Fig.~\ref{fig:basicidea}. 
The resulting spectrogram allows for direct identification of the group delay of the attosecond pulse, as shown by the white dashed line in the two lower panels of Fig.~\ref{fig:basicidea} for a Fourier limited (left lower panel) and linearly chirped (right lower panel) pulse, respectively. 
In general, the quantum beating is of the generic form 
\begin{align}
P(\epsilon,\tau) = A(\epsilon) + B(\epsilon) \cos\{\Delta\omega[\tau+\tX(\omega)]\},  
\label{genericP}
\end{align}
where $\Delta\omega=\Delta\epsilon/\hbar$ is the angular frequency spacing of the bound wave packet and $\tX(\omega)$ is the attosecond pulse group delay as a function of angular frequency, $\omega$. The mapping between pulse frequency and electron energy is given by the photoelectric effect, 
\begin{align}
\epsilon=\hbar\omega-I_p^{(wpk)}, 
\end{align}
where $I_p^{(wpk)}=|\epsilon_1+\epsilon_2|/2$ is the effective binding energy of the wave packet and the bound state energies are $\epsilon_{j=1,2}<0$. The generic form of the photoelectron spectrogram, $P(\epsilon,\tau)$, is the main result of the PANDA method and it will be justified in Sec.~\ref{sec:derivation}. Similar to the SPIDER technique \cite{Iaconis1998}, the inner workings of PANDA are based on spectral-shearing interferometry. In contrast to SPIDER, however, a full photoelectron spectrogram $P(\epsilon,\tau)$ over both $\epsilon$ and $\tau$ must be determined for the PANDA method.  

The main advantage of the PANDA method is that it does not involve any atomic latency effects that could obscure the reconstruction procedure, see Sec.~\ref{sec:comment}. In this sense the PANDA method is a cleaner method than characterization methods based on laser-assisted photoionization that ``suffer'' from atomic latency due to  Wigner delays and long-range laser-Coulomb interactions, as implied by Eq.~(\ref{tauA}). Finally, in Sec.~\ref{sec:angleresolved}, we will show that extending the PANDA method to angle-resolved photoelectrons allows for studies of energy- and angle-dependent delays from time-dependent bound wave packets on the time scale of few attoseconds that can be measured relative to latency-free (angle-integrated) PANDA.   

\subsection{Derivation of PANDA}
\label{sec:derivation}
Consider an atom prepared in a coherent superposition of two bound states as illustrated in Fig.~\ref{fig:basicidea}~(b),   
\begin{align}
\ket{\psi(t)} &=
\sum_{j=1,2} c_j \ket{j} \exp[-i\epsilon_j t/\hbar]
%\nonumber \\
% &=\left\{
%c_1\ket{1}+c_2\ket{2}\exp[-i\Delta\epsilon t/ \hbar]
%\right\}\exp[-i\epsilon_1 t/\hbar],
\label{boundwavepacket}
\end{align}
where $c_j>0$ are constant amplitudes that describe the wave packet states with quantum numbers, $j\in\{1,2\}$, and non-degenerate energies, $\epsilon_j$. The use of constant complex amplitudes in the wave packet is an idealization that assumes (i) that the preparation and ionization steps are sequential and (ii) that the depletion of the wave packet due to decay or ionization can be neglected. 
A test pulse with a controllable delay, $\tau$, in the time domain can be expressed in terms of the Fourier components of an unshifted pulse by use of the shift theorem,  
\begin{align}
&\tilde \EX(t,\tau)=\tilde \EX(t-\tau,0) 
\nonumber  \\
&=\frac{1}{2\pi}\int d\omega |\EX(\omega)|
\exp[-i\omega(t-\tau)+i\phiX(\omega)],
\label{delayedpulse}
\end{align} 
where $|\EX(\omega)|$ and $\phiX(\omega)$ are the spectral amplitude and phase of the test pulse with $\tau=0$. The control delay, $\tau$, of the test pulse can be seen as an experimental parameter that shifts the pulse in time as a perfect delay stage.  
%Spatial propagation of the pulse or inter-atomic interactions will not be considered. 
%

For simplicity, we consider a central-field model with eigenstates $\ket{j}$, corresponding to quantum numbers: $(n_j,\ell_j,m_j,\sigma_j)$. These wave packet states must have the same parity and they should satisfy selection rules that allow one-photon ionization by the test pulse to reach a common final state consisting of both ion and photoelectron. The complex amplitude for an electron to reach a final continuum state, $\ket{f}$ with quantum numbers $(\epsilon_f,\ell_f,m_f,\sigma_f)$, by photoionization from the bound wave packet, is given by (here in length gauge)
\begin{align}
a_f(t)=&
-\frac{e}{i\hbar}\int_{-\infty}^{t}dt' 
\tilde \EX(t,\tau)
\braOket{f}{z}{\psi(t)}\exp[i\epsilon_f t/\hbar]
\nonumber \\
=&
-\frac{e}{i\hbar} 
\sum_{j=1,2}
\EX(\omega_{fj}) 
\braOket{f}{z}{j}
c_j \exp[\omega_{fj}\tau],
\end{align}
where $z$ is the operator of a linearly polarized test pulse within the dipole-approximation with well-known selection rules for the angular momentum:  $\ell_f=|\ell_j \pm 1|$, $m_f=m_j$, and spin, $\sigma_f=\sigma_j$. Note that the angular frequencies absorbed from the test pulse in the transitions from the bound wave packet, $\omega_{fj}=(\epsilon_f-\epsilon_j)/\hbar$, are different due to energy conservation. This is the key to spectral-shearing interferometry in the photoelectron spectrum, where different spectral amplitudes of the test pulse can be made to interfere. By squaring the complex amplitudes of the photoelectrons we obtain the final probability density resolved over photoelectron energy and delay, 
\begin{align}
P(\epsilon,\tau) = 
A_1 + A_2 + B\cos\Theta
\label{Pofepstau}
\end{align}
where the direct ionization terms from state $j=1,2$ are
\begin{align}
A_{j} = 
\left(\frac{e}{\hbar}
|\EX(\omega_{fj})| \braOket{f}{z}{j} c_j \right)^2, 
\label{A}
\end{align}
the ionization cross-term magnitude is
\begin{align}
B=\frac{2e^2}{\hbar^2}
\braOket{f}{z}{1}\braOket{f}{z}{2}
|\EX(\omega_{f1})E(\omega_{f2})|c_1c_2 
\label{B}
\end{align}
while the ionization cross-term phase is
\begin{align}
\Theta = \omega_{21}\tau + 
\phi_X(\omega_{f1})-\phi_X(\omega_{f2}). 
\end{align}
In writing these equations we have made use of the assumption that the complex amplitudes of the wave packet are positive and that the dipole matrix elements are real. Provided that the wave packet is suitably designed, the variation of the spectral phase is small over the frequency splitting of the wave packet, so that the phase difference can be rewritten as level splitting time group delay, 
\begin{align}
\phi_X(\omega_{f1})-\phi_X(\omega_{f2}) \approx 
\Delta\omega  \tau_{X}^{(GD)}(\omega),
\end{align}
where $\omega = (\omega_{f1}+\omega_{f2})/2$ is the mean angular frequency to reach the final state from the wave packet and $\Delta\omega=\omega_{f1}-\omega_{f2}>0$ is the angular frequency splitting of the wave packet. In this way the generic form of a PANDA spectrogram shown in Eq.~(\ref{genericP}) is recovered.

\subsection{Comments on dipole matrix elements}
\label{sec:comment}
One question that may arise is why we can assume that the dipole matrix elements are {\it real} --  are dipole matrix elements to the continuum not {\it complex} quantities in general? So far we have only discussed dipole transition to well-defined angular momentum states within a central field model, where both bound and continuum radial wavefunctions can be chosen to be real functions \cite{Friedrich2006}. This implies that any such bound-to-continuum matrix element of a real operator, e.g. $z$,  will be real as well. In contrast, the photoionization amplitude to a final momentum state, with a given direction $\mathbf{\hat k}$, is complex because the final state is given by
\begin{align}
\psi^-_{\mathbf{k}} = \frac{1}{k^{1/2}}\sum_{L=0}^{\infty}\sum_{M=-L}^{L}
i^L e^{-i\eta_L} Y_{LM}^*(\mathbf{\hat k}) Y_{LM}(\mathbf{\hat r})R_{\epsilon L}(r), 
\label{eq:kstate}
\end{align}
where $\eta_L$ are scattering phases and $R_{\epsilon L}(r)$ are energy-normalized radial wavefunctions, as denoted in our earlier work \cite{PabstPRA2016}. 
For this reason the PANDA will show latency in angle-resolved detection, but {\it not} in angle-integrated detection. This point will be demonstrated by numerical calculations in Sec.~\ref{sec:angleresolved}.

{\it Cooper minima} occur when radial dipole matrix elements go to zero and change sign from negative to positive \cite{cooper:1962}. One such minimum is associated with a shift of the PANDA signal by a half quantum beat period according to Eq.~(\ref{Pofepstau}) as the sign of $B$ changes in Eq.~(\ref{B}). Does this cause problems for PANDA? We do not believe that it is a problem if bound wave packets of equal angular momentum $\ell_0$ are used. 
We have found that transitions to the continuum from different principal quantum numbers, say $n_0$ and $n_0+1$ with the same $\ell_0$, tend to have closely placed Cooper minima in kinetic energy. Now, if {\it both} dipole matrix elements in $B$ change sign at (almost) the same kinetic energy, the shift of the quantum beat is canceled and the PANDA signal will not be affected.   
We interpret this effect as due to the fact that bound states with the same angular momentum behave similarly close to the core due to the dominating centrifugal potential \cite{Friedrich2006}.%, which means that the condition for a Cooper minimum will be satisfied at almost the same photoelectron energy.  

{\it Fano resonances} are examples of atomic phenomena that go beyond the central-field model due to electron--electron correlation \cite{fanoPR1961}. Using Fano's theory \cite{ZhaoPRA2005}, in the  case the of a single continuum coupled to a single resonance, the {\it correlated} dipole matrix element, $Z_{fj}$, can be written in terms of the real {\it uncorrelated} matrix element, $z_{fj}$, as 
\begin{align}
Z_{fj}=\frac{(q_j+\epsilon_F)}{(1-i\epsilon_F)}z_{fj},
\label{fano}
\end{align}
where $\epsilon_F=(\epsilon-\epsilon_{r})/(\Gamma/2)$ is the photoelectron energy shifted by the resonance energy, $\epsilon_r$, and rescaled by the inverse resonance lifetime, $\Gamma$. The correlated dipole matrix element is clearly complex and the question arises if this can affect the PANDA method? We do not think that the PANDA signal is affected by such a resonance because it is the phase difference from two different wave packet states to the {\it same} final continuum state that may affect the PANDA measurement. This phase difference corresponds to the argument of the ratio of two correlated Fano dipole matrix elements,  
\begin{align}
\arg\left(\frac{Z_{fj}}{Z_{fj'}}\right) = \arg\left(\frac{(q_j+\epsilon_F)}{(q_{j'}+\epsilon_F)}\right) = 0 \, \mathrm{or} \, \pi,
\end{align}
where we find that the complex denominators have canceled in the first step. In the second step we assume that the uncorrelated matrix elements and $q$-factors are real. The reason for this cancellation of the Fano phases is that the phase variation is determined solely by the inverse lifetime of the resonance, which does not depend on the initial wave packet state $\ket{j}$. In our earlier work, we have showed numerically that such a resonance does not affect the PANDA technique, by performing simulations within the Time-Dependent Configuration Interaction Singles (TDCIS) for the case of neon atoms with low $2s^{-1}ns$ autoionizing states \cite{PabstPRA2016}. By extension, a resonance coupled to several continua should not affect PANDA either, because it is possible to perform a transformation back to the case where only one continuum couples to the resonance \cite{fanoPR1961}.   

{\it Inner-shell photoionization and shake-up processes} has been estimated in our previous work within the Hartree-Slater formalism \cite{PabstJPB2017}. Clearly, inner-shell photoionization can easily overwhelm the signal from photoionization of the bound outer wave packet, but we found that inner-shell photoionization does not affect the probability modulations of PANDA as long as the {\it independent-particle approximation} is valid. Effects {\it beyond} independent particles {\it can} manifest itself as new probability modulations in PANDA. For instance, electron correlation effects that are associated with inner-shell ionization can stimulate transitions in the outer bound wave packet by so-called shake-up effects. It would be interesting to study these aspects of electron correlation further, but this remains beyond the scope of this tutorial. 

\subsection{Design of bound wave packet}
\label{sec:designwp}
The time-dependent bound wave packet used for PANDA measurements must satisfy three basic compatibility requirements with the test pulse: (i) the central angular frequency must exceed the effective wave packet binding energy to ensure ionization, $\omega_X \gg I_p^{(wpk)}$, (ii) the test pulse bandwidth must be larger than the wave packet energy splitting to ensure spectral shearing phenomena, $\Delta\omega_X\gg\Delta\omega$, and (iii) the lifetime of the bound wave packet must be much longer than the test pulse duration, $1/\Gamma^{(wkp)} \gg \Delta \tau_\mathrm{X}$. For simplicity, we will only discuss idealized bound wave packets with infinite lifetime, $\Gamma^{(wpk)}=0$.
%, but dephasing and coherence studies of bound wave packets is a possible for future direction PANDA-like experiments. 
%
\begin{figure}[!htb]
\includegraphics[width=0.45\textwidth]{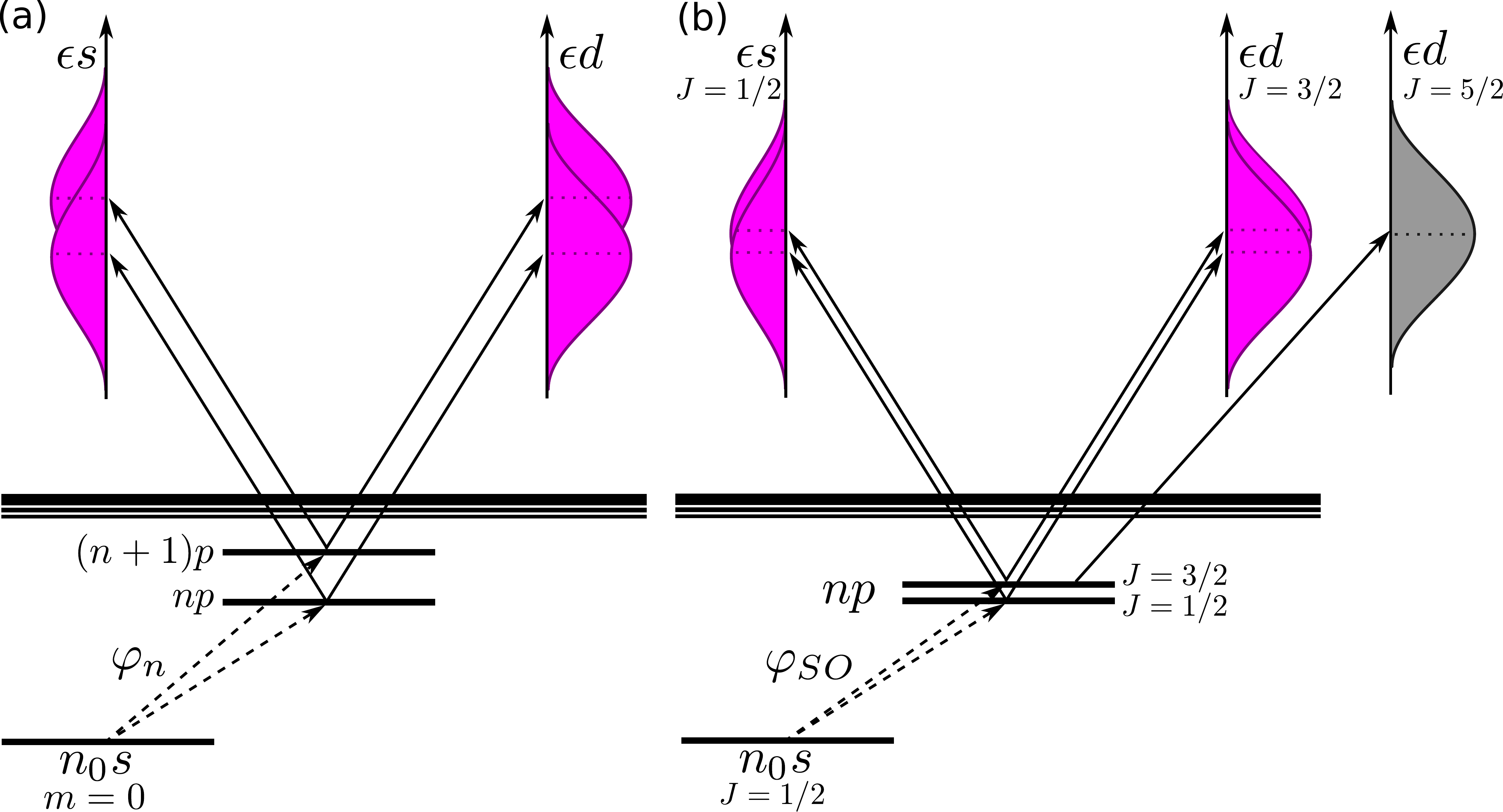}
\caption{
Level schemes for PANDA method in alkali atom (H like) with (a) nearest principal quantum number and (b) spin-orbit wave packets (see Appendix~\ref{sec:so}).
\label{fig:levelalkali}}
\end{figure}
Clearly, there are many different atomic systems and excitations to consider for the PANDA method. In general, the photoionization cross sections of atomic orbitals decrease rapidly with both kinetic energy of the photoelectron and initial principal quantum number. 
Alkali atoms have the advantage that the excitation energies from the ground state are in optical/UV regime, which makes coherent state preparation readily feasible by short laser pulses. In the Sec.~\ref{sec:angleresolved}, we will consider alkali atoms with the simplest possible sequential excitation scheme, illustrated in Fig.~\ref{fig:levelalkali}~(a) corresponding to nearest principal quantum number wave packets. In Appendix~\ref{sec:so} we also discuss the case of spin-orbit (SO) wave packets in alkali atoms illustated in Fig.~\ref{fig:levelalkali}~(b). 
Other possible schemes include inner-shell photoionization with shake-up \cite{PabstJPB2017} and photoionization of spin-orbit hole wave packets, similar to the experiment on coherently excited Krypton ions that were created by strong-field ionization by Gouliemakis {\it et al.} in 2010 \cite{GouliemakisNature2010}.  
\begin{figure}[!htb]
\includegraphics[width=0.45\textwidth]{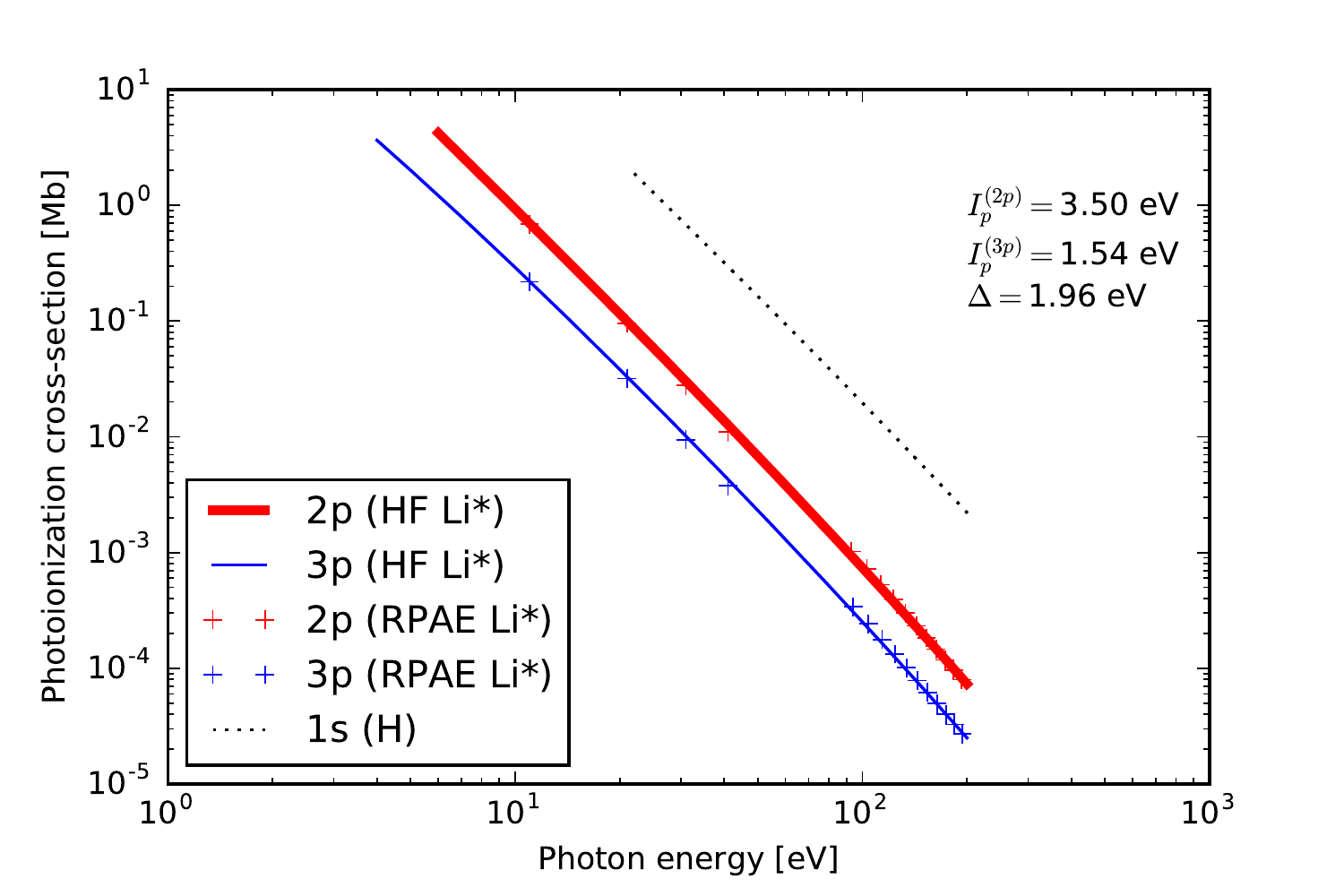}
\includegraphics[width=0.45\textwidth]{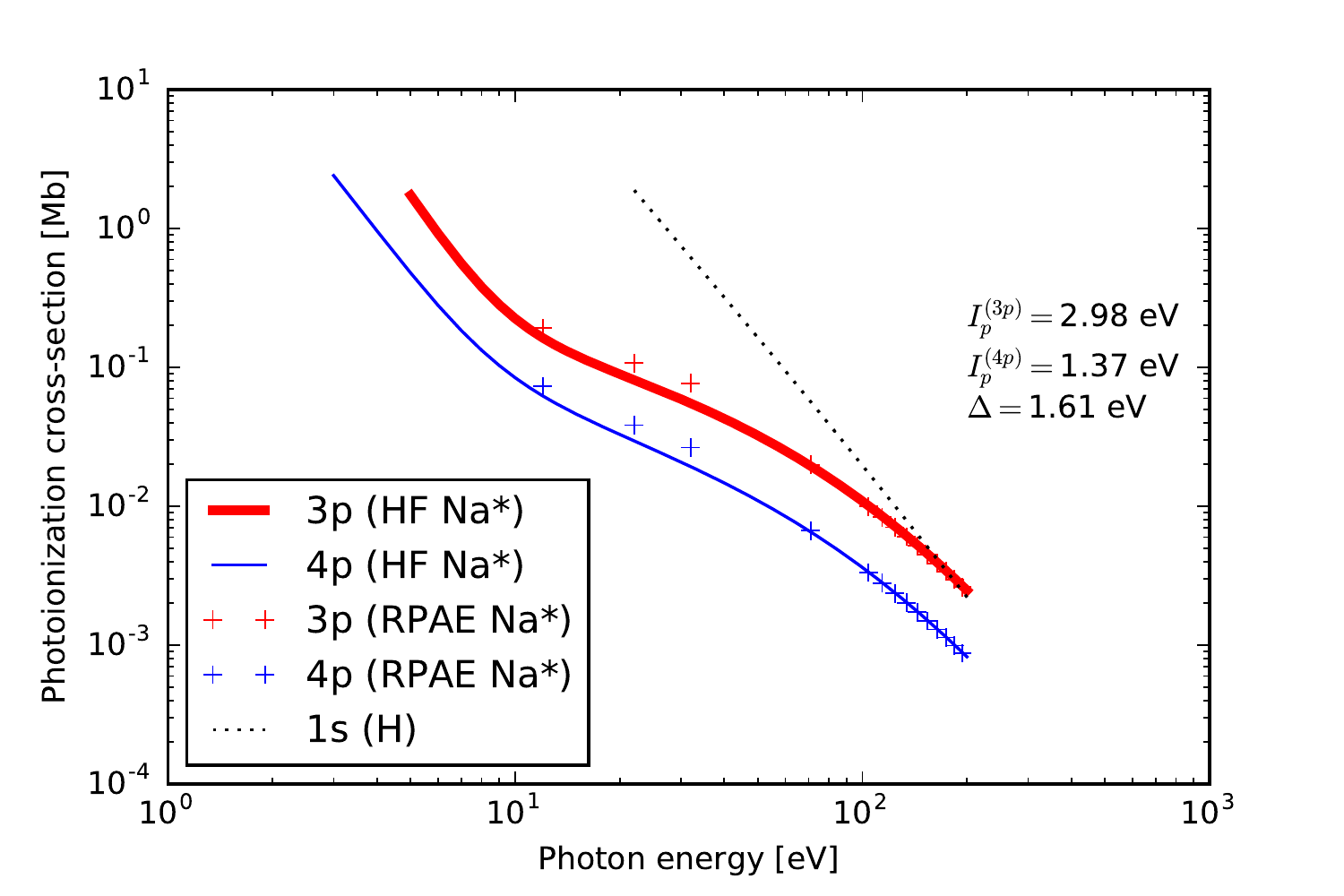}
\includegraphics[width=0.45\textwidth]{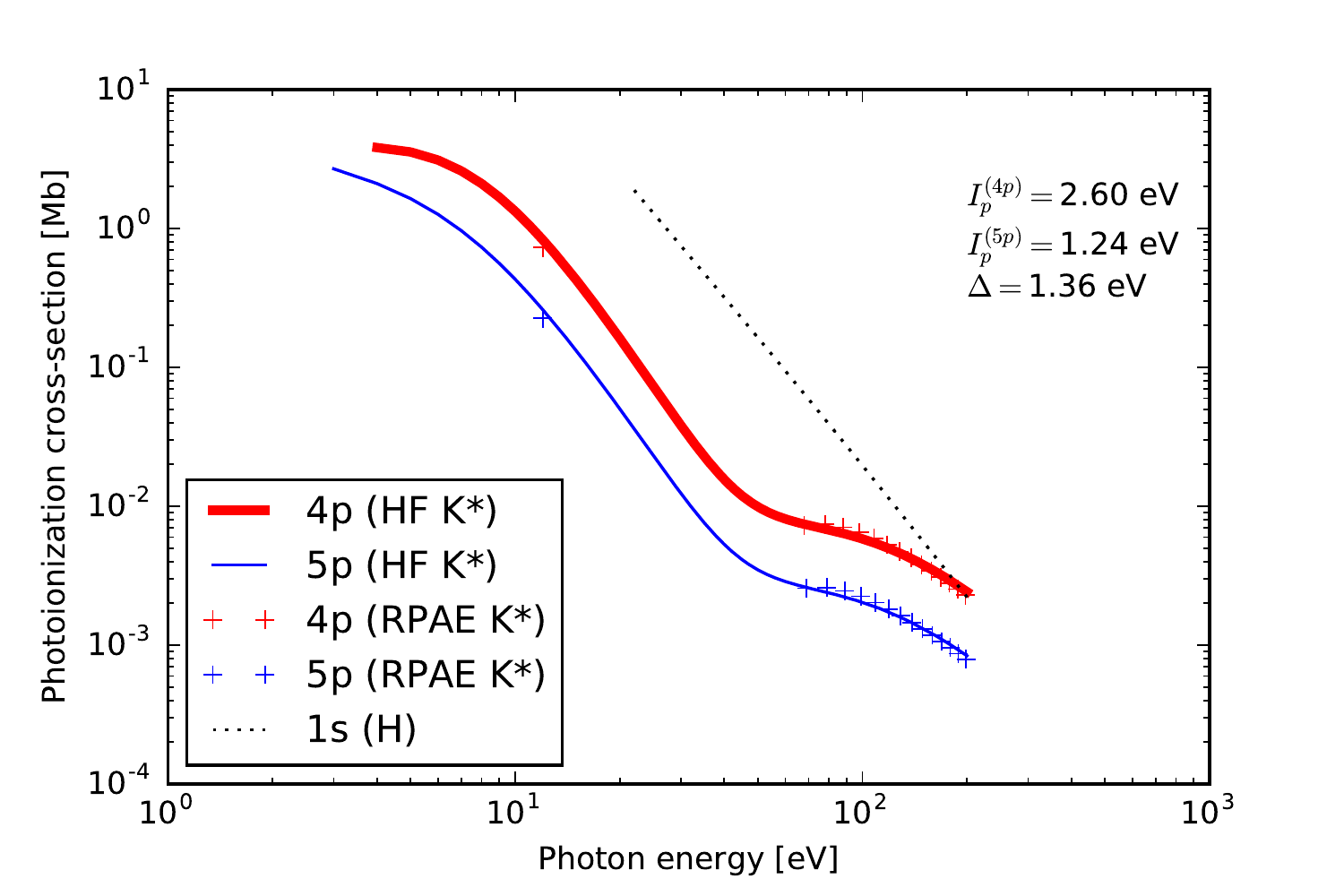}
\caption{
Photoionization cross sections from excited $np$ states of alkali atoms lithium (Li$^*$), sodium (Na$^*$) and potassium (K$^*$) compared to the exact cross section for photoionization from the $1s$ ground state of hydrogen (H). The Hartree-Fock (HF) results compare well with the Random Phase Approximation with Exchange (RPAE) in regions far from thresholds of the core.   
\label{fig:CSalkali}}
\end{figure}

\subsection{Numerical results for PANDA in alkali atoms}
\label{sec:angleresolved}
In Fig.~\ref{fig:CSalkali} we show the photoionization cross sections from the lowest excited states of odd parity for alkali atoms Li, Na and K. The calculation is performed using the cross section formula for linearly polarized light (here length gauge)
\begin{align}
\sigma(\omega) = 4\pi^2\alpha\omega\sum_{\ell_f}|\braOket{f}{z}{j}|^2\times a_0^2[Mb], 
\end{align}
where $\alpha \approx 1/137$ is the fine structure constant, $\omega$ and $\braOket{f}{z}{j}$ are the photon energy and dipole matrix elements in atomic units, while $a_0^2[Mb]=28.0028$ is the numerical conversion factor from atomic units to megabarn (1 \,Mb = $10^{-22}$\,m$^2$). 
The dipole matrix elements are computed using the independent-particle Hartree-Fock approximation (full lines). Correlation effects due to the atomic core are found to be small using the Random Phase Approximation with Exchange (RPAE) in energetic regions away from thresholds of the core (plus signs) \cite{Pabst2016}. 
The exact photoionization cross section from the ground state of hydrogen is shown for comparison (dotted line). 
At high photon energy the hydrogen cross-section follows a $\omega^{-7/2}$ scaling law and photoionization from excited states of H have a $n^{-3}$ scaling of the cross sections \cite{bethe-salpeter}. 
Our results for Li* show that the cross section from the lowest excited state is lower by one order of magnitude compared to H. The cross sections for excited states of Na and K are structured because of the presence of Cooper minima in the $n_0p\rightarrow \epsilon_fd$ transition. The cross sections of Na* and K*  are larger than that of Li* at high energy and they are comparable to that of the ground state of H at 200 eV photon excitation.   
%The high energy scaling of Na and K does not appear to be as sharp as for Li. 
The ratio of nearest principal quantum number cross sections is $\sim 1/3$ between all alkali atoms in our study at high energy. This compares well with the cross section ratio expected for excited hydrogen, $(n+1)^{-3}/(n)^{-3}$, with the lowest excited states, $3^{-3}/2^{-3} \approx 0.3$. As the principal quantum number $n$ increases the cross section ratio is expected to reach unity. The energy difference between nearest principal quantum number states scales as $1/n^3$, which implies that the quantum beating period of the wave packet can be chosen to be extremely long $\sim n^3$.   

\begin{figure}[!htb]
\includegraphics[width=0.5\textwidth]{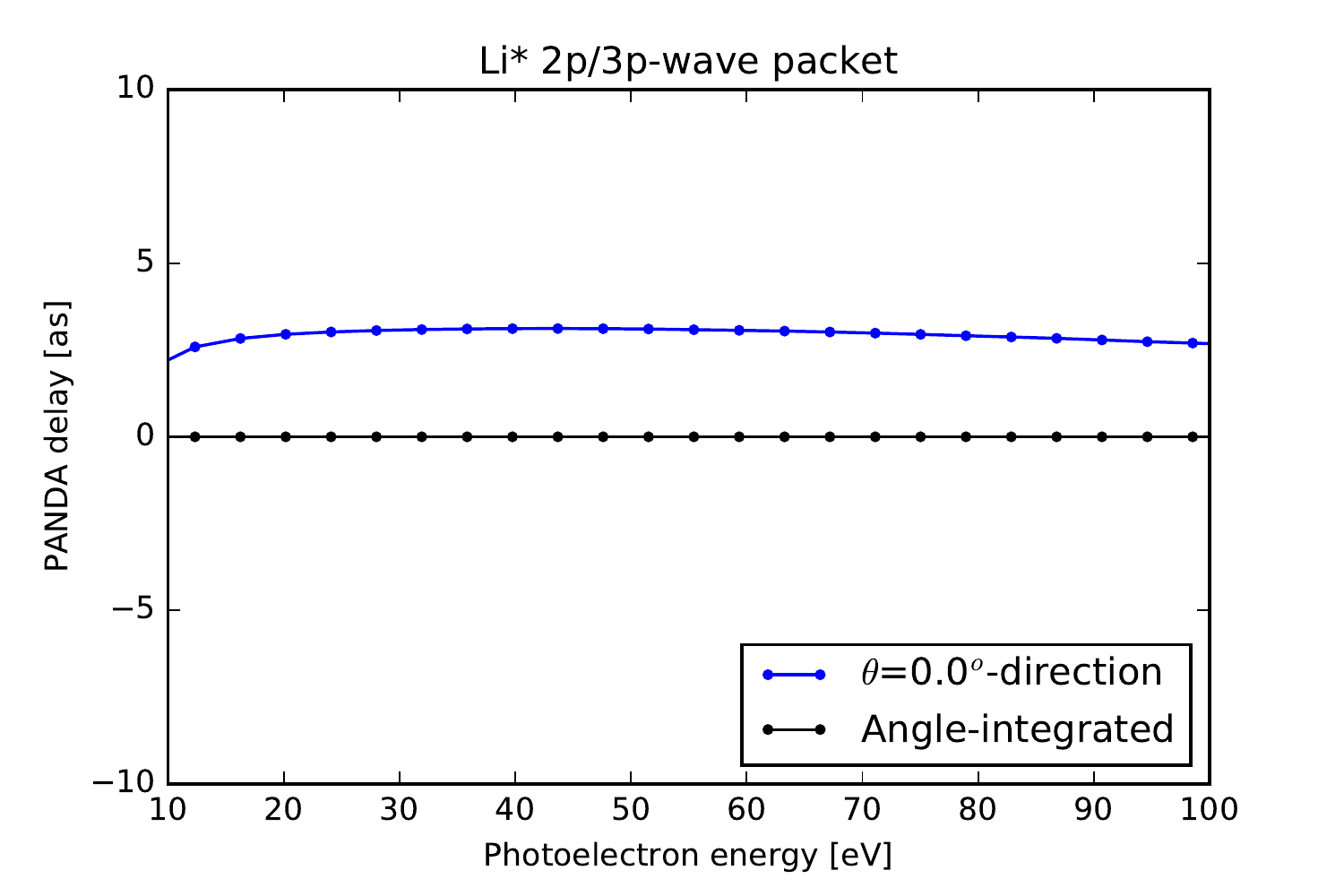}
\includegraphics[width=0.5\textwidth]{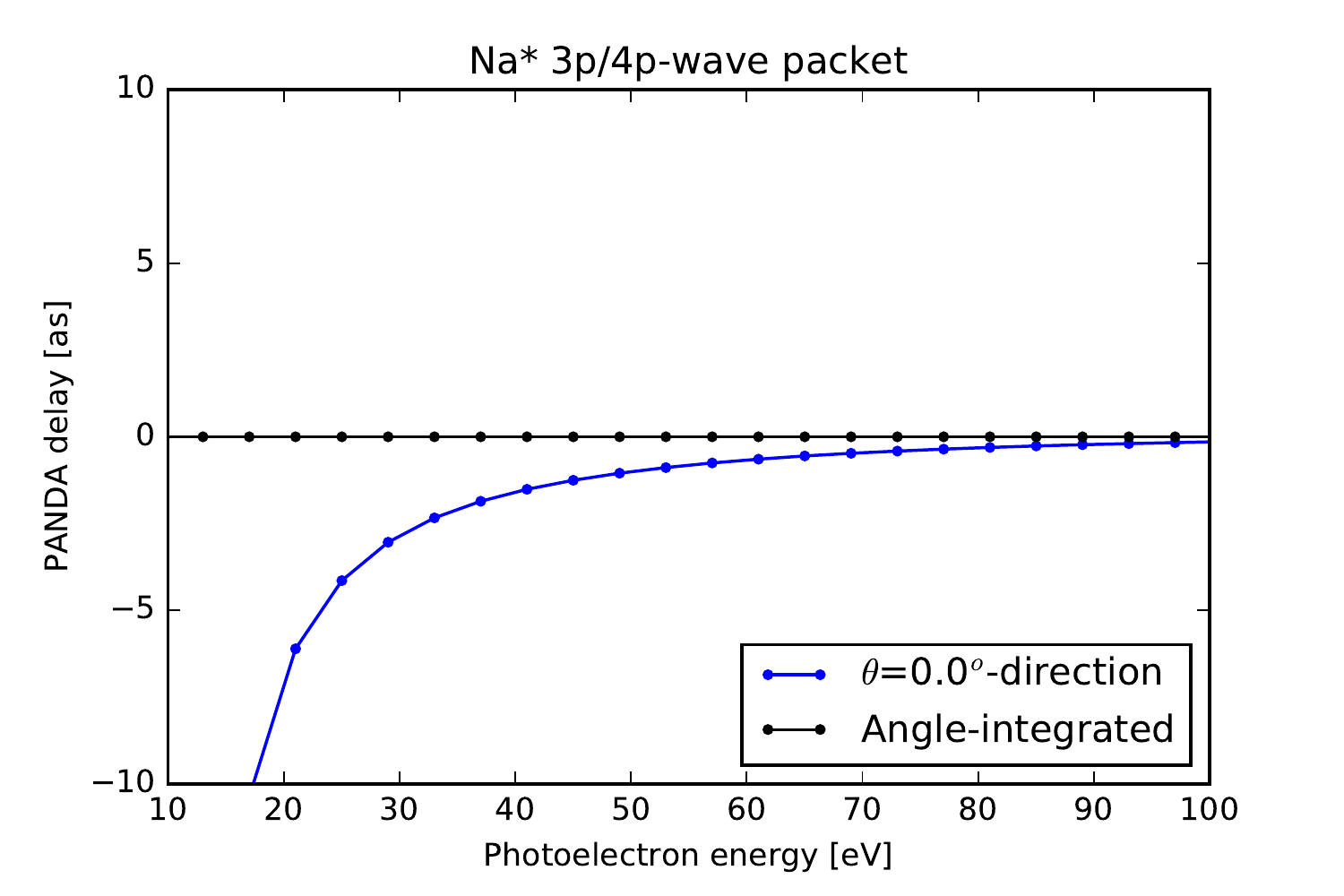}
\includegraphics[width=0.5\textwidth]{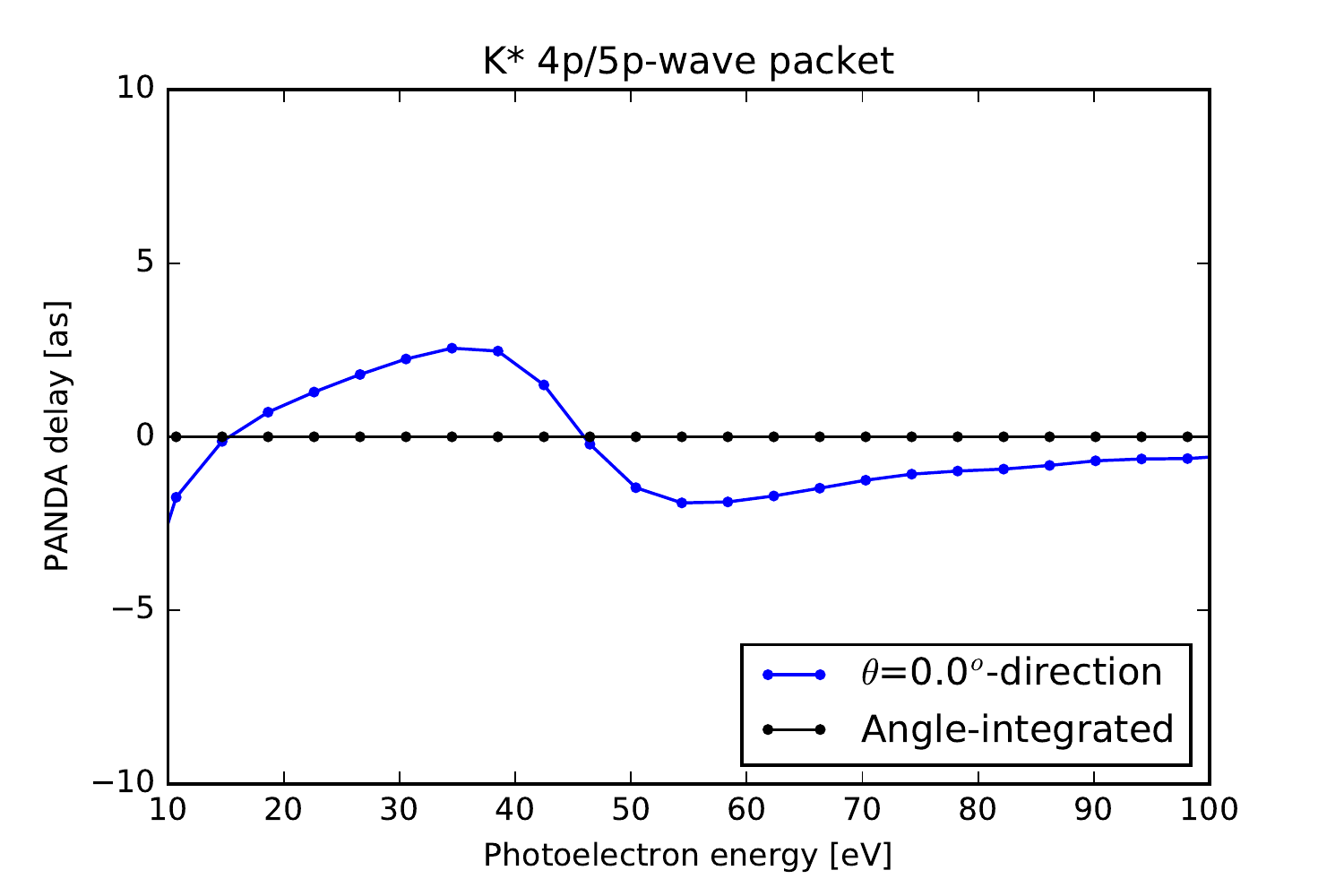}
\caption{
Delays of quantum beats for Li*, Na* and K* with angle-integrated and angle-resolved emission along the test pulse polarization direction ($\theta=0$) with  principal quantum numbers wave packets as explained the main text. 
\label{fig:delayresolved}}
\end{figure}

Next, we show that it is important that the photoelectron is integrated over {\it all} emission angles in order to obtain a PANDA result that is {\it free} from atomic latency. The angle-resolved PANDA delays are computed using a complex final state for photoemission in a given direction within the Hartree-Fock approximation using Eq.~(\ref{eq:kstate}), as explained in Ref.~\cite{Pabst2016}. In Fig.~\ref{fig:delayresolved} we compare angle-integrated photoelectron emission with photoelectron emission along the polarization of the test pulse for wave packets in Li*, Na* and K* (all excited wave packet states are $p$-waves, which implies that the reached continuum consists of $s$- and $d$-waves).  
While there is zero PANDA delay for the angle-integrated cases, the angle-resolved measurements show vastly different delays for all atoms. The Li* result shows a positive delay of $\sim 3$\,as over the entire energy region, while the Na* result instead shows a negative delay that increases in magnitude with decreasing photoelectron energy. The K* result shows a structured delay that goes from being positive to negative close to 47\,eV \cite{PabstPRA2016}. This structure is attributed to the presence of Cooper minima in the photoionization cross section of the excited states in K* at $\sim 42$\,eV in the $np\rightarrow\epsilon d$ channel. Similarly the Na* structure can be explained by Cooper minima at lower kinetic energies $\sim 7$\,eV. In contrast we find no Cooper minima in Li*, which suggests the positive and featureless delay.

\begin{figure}[!htb]
\includegraphics[width=0.5\textwidth]{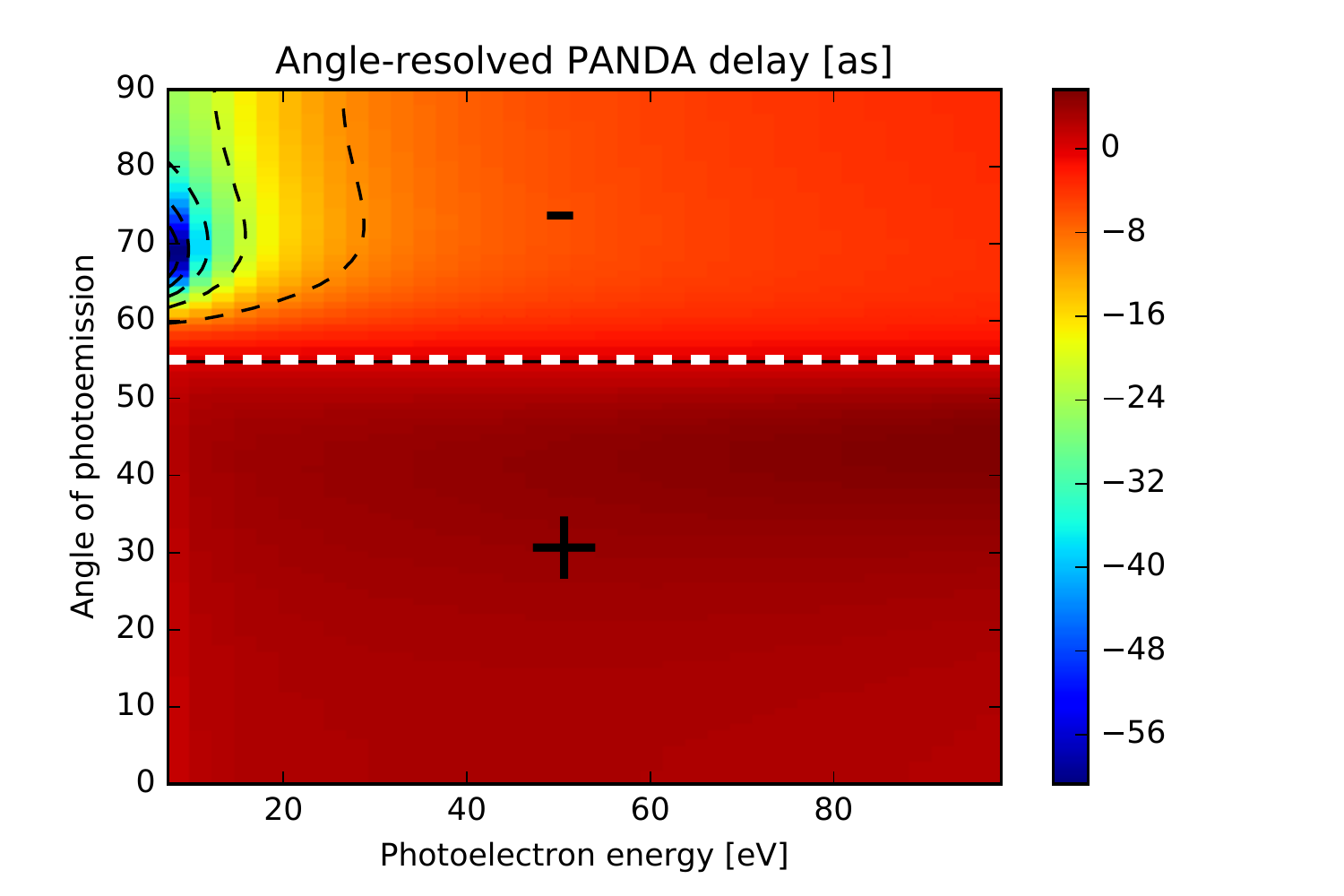}
\includegraphics[width=0.5\textwidth]{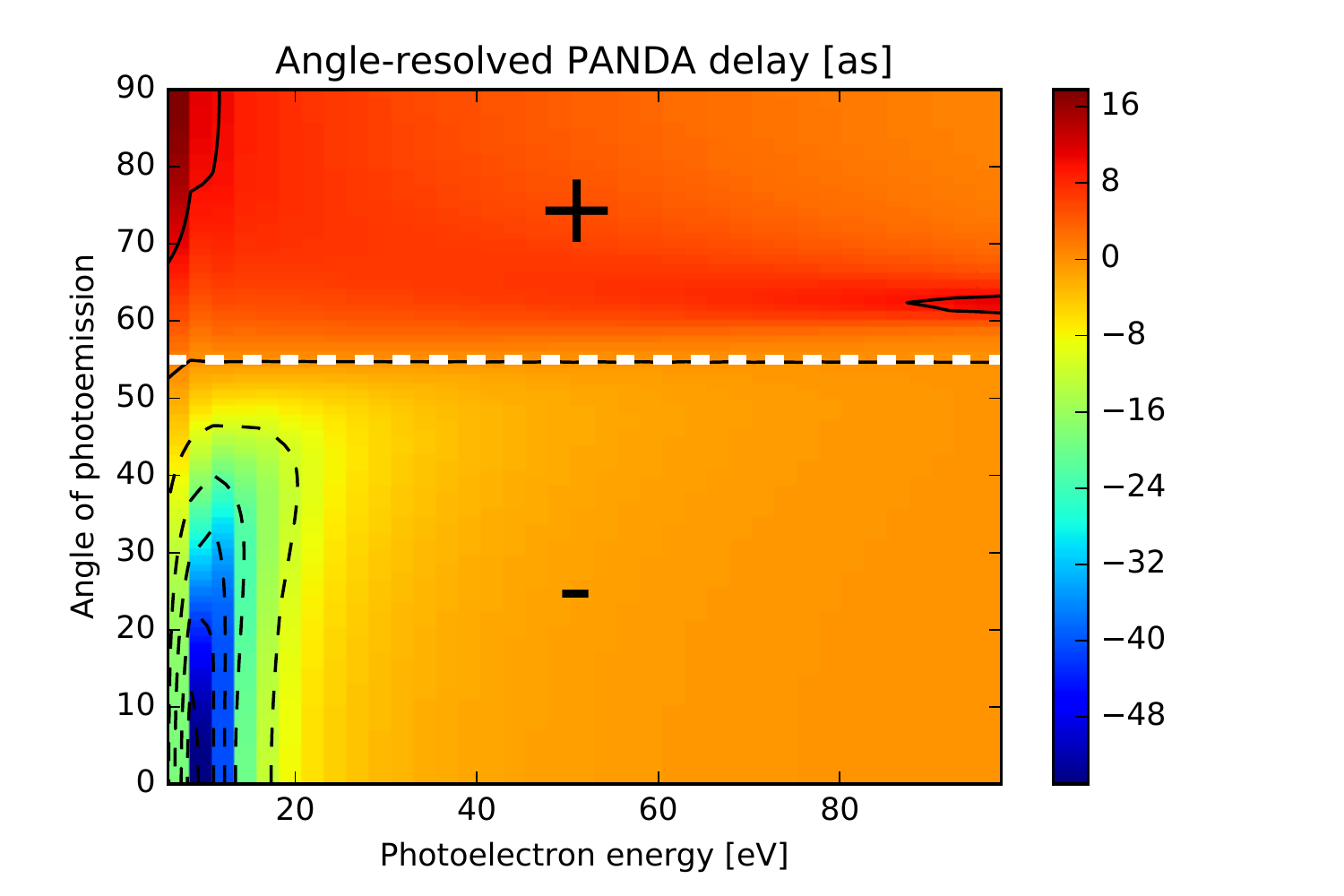}
\includegraphics[width=0.5\textwidth]{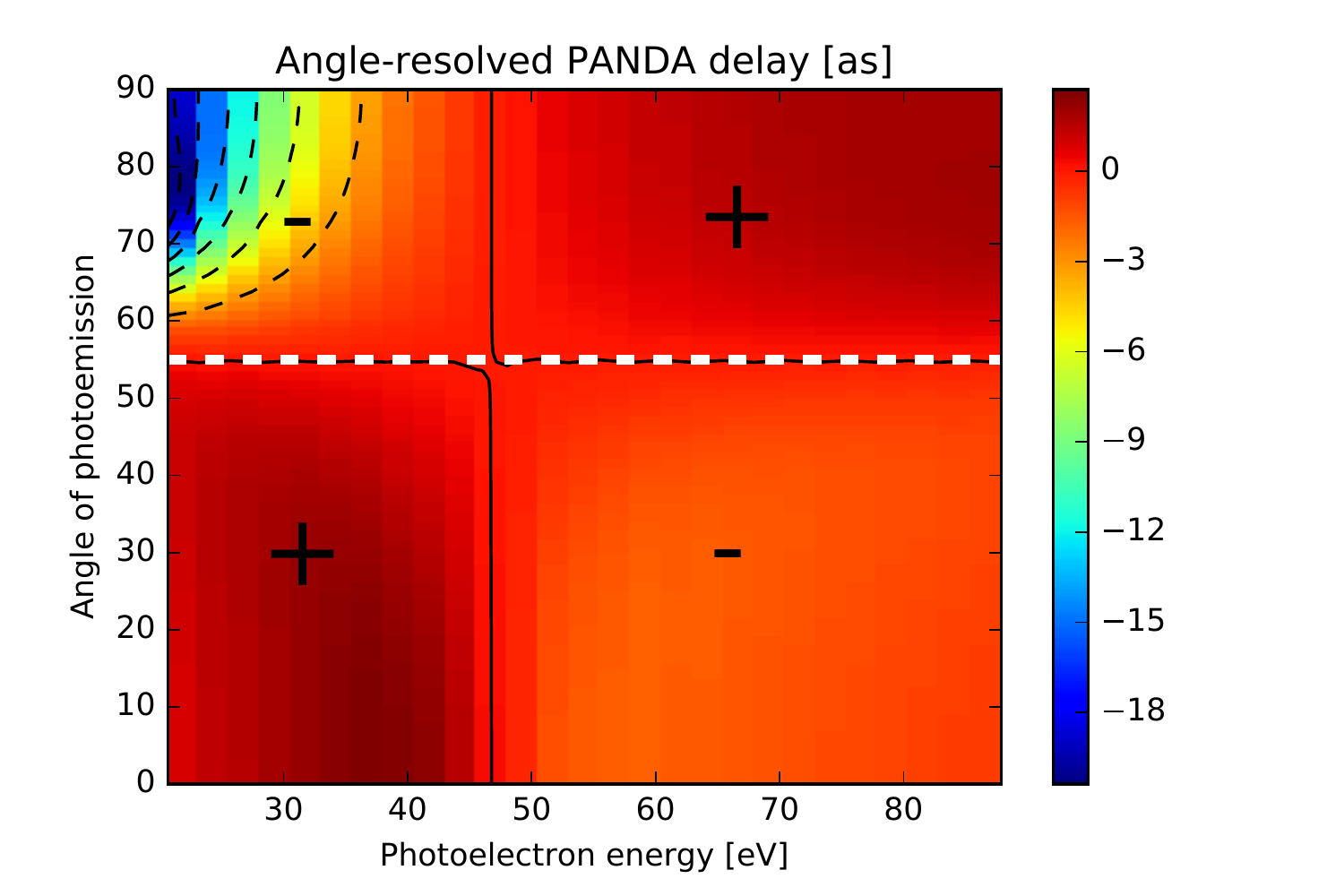}
\caption{
Delays of quantum beats from wave packets in Li*, Na* and K* resolved over azimuth angle and photoelectron energy with initial nearest principal quantum numbers wave packets of p-waves. Plus and minus signs indicate areas of positive and negative delay for clarity.  
\label{fig:delaymap}}
\end{figure}
In order to show how the angle-integrated PANDA delay can ``sum up to zero'' in Fig.~\ref{fig:delayresolved}, we show in Fig.~\ref{fig:delaymap} the angle-resolved delay as a function of the azimuth angles relative to the test pulse polarization for Li*, Na* and K*.  
In all the atoms we find a universal zero delay for all energies that occurs at $\sim 55^o$ where the partial $d$-wave vanishes and only the isotropic $s$-wave remains. On the other side, $\theta>55^o$, the delay changes sign. Clearly, it is this sign change (marked by $+$ and $-$ in Fig.~\ref{fig:delaymap}) what allows for the angle-integrated result to sum up to zero. 

In some regions the angle-resolved delay becomes quite large (tens of attoseconds) to compensate for the fact that the emission probability density is low in that direction. In Li* this happens close the threshold where the $s$- and $d$-waves interfere destructively at large emission angles. In Na* the change of PANDA delay for $\theta>55^o$ is explained by the fact that the Cooper minima for angle-resolved emission moves to much lower kinetic energy (below the separated $d$-wave result) when the sign of the $d$-wave changes at $\theta\approx 55^o$. A similar effect occurs in K*, where for $\theta>55^o$ only the negative delay region above the angle-resolved Cooper minima is observed. 

In general we find that the angle-resolved delay from coherent bound wave packets is changing on the order of tens of attoseconds in the vicinity of Cooper minima. Experimental measurements of these effects will be challenging due to low photoionization cross sections, but should in principle be possible by performing {\it relative} delay measurements between angle-resolved and angle-integrated photoelectrons.   

\section{Summary and outlook}
In this tutorial we have provided a short review of attosecond pulse characterization techniques. We focused on the recently proposed Pulse Analysis by Delayed Absorption (PANDA) method \cite{PabstPRA2016}, and discussed its various possible implementations and advantages. 
The main merit of PANDA is that it is a latency free pulse characterization method that is insensitive to scattering phase shifts and atomic resonances. Combining the PANDA method with angle-resolved photoelectron detection should allow for measurement of delays in photoionization from bound wave packets on the order of tens of attoseconds.  

In closing, we have recently attempted to extend PANDA to attosecond transient absorption, where transmitted photons are detected instead of angle-resolved photoelectrons \cite{DahlstromJO2017}. In this case, we found that it was {\it not} possible to extract the exact attosecond pulse shape because the susceptibility of the process was complex due to scattering of  photons and electrons in the continuum above the ionization threshold.    

\begin{acknowledgments} 
J.M.D. is funded by the Swedish Research Council, Grant No. 2014-3724. 
S.P. is funded by the NSF through a grant to ITAMP.
E.L. is funded by the Swedish Research Council, Grant No. 2016-03789. 
J.M.D and E.L. acknowledge funding from the Knut and Alice Wallenberg Foundation.   
\end{acknowledgments}

\appendix 

\section{PANDA with spin-orbit wave packet}
\label{sec:so}
In this appendix we consider spin-orbit (SO) wave packets for the PANDA method consisting of a {\it single active electron} in alkali atoms. Specifically, we consider the K atom which is sequentially excited from its ground state, $4s$, to its excited state, $4p$, by a laser pulse, $\EL$. The spectral support of $\EL$ is such that it excites both fine-structure levels of the excited state, $4p_{j'=1/2}$ and $4p_{j'=3/2}$,  that will be denoted by their total angular momentum $j'$ for brevity in the following. The corresponding SO wave packet evolves freely as  
\begin{align}
\ket{\psi^{(1)}(t)}=\sum_{j'=1/2,3/2}\cj\exp[-i\omega_{j'} t]\ket{4p_{j'}m}, 
\label{SOwp}
\end{align}
where the complex amplitude $\cj$ can be approximated by first order perturbation theory as   
\begin{align}
\cj=-\frac{e}{i\hbar}z_{j'i}\EL(\omega_{j'i}).
\label{cj}
\end{align}
In Eq.~(\ref{cj}), $z_{j'i}=\braOket{4p_{j'}m}{z}{4s_{1/2}m}$ denotes the dipole matrix elements between the initial state $4s_{1/2}m$ (denoted $i$) and the excited states $j'$. The total magnetic quantum number $m=\pm1/2$ is conserved in all interactions due to the linear polarization of the fields. 
The angular frequencies, $\omega_{j'i}=\omega_{j'}-\omega_i$, correspond to the excitation energies $1.610$\,eV  for $j'=1/2$ and $1.617$\,eV for $j'=3/2$ relative to the initial state energy, $\epsilon_i=\hbar\omega_i=-4.341$\,eV \cite{NIST:database}. The time scale for the SO dynamics is determined by the inverse splitting of the excited levels, $\ESO=\hbar\wSO=7.15517$\,meV, which translates to a period on the sub-ps time scale, $\TSO=2\pi/\wSO=577.998$\,fs. 
% ESO=1.6171129384-1.6099577731=0.007155165
% TSO=4.135667662E-15/0.007155165=577.997525144

Next, the test pulse, $\EX$, is used to photoionize the excited K atom so that an electron is promoted to continuum state with energy $\epsilon>0$. The dipole allowed continuum states include $\epsilon s_{j=1/2}$, $\epsilon d_{j=3/2}$ and $\epsilon d_{j=5/2}$ (denoted by their total angular momentum $j$ for brevity).  After the interaction with $\EX$ has ceased the complex amplitude for the final state $j$ is 
\begin{align}
\ck=\left(\frac{-e}{i\hbar}\right)^2\sum_{j'=1/2,3/2}\EX(\omega_{jj'})\EL(\omega_{j'i})z_{jj'}z_{j'i},
\end{align}
where the second dipole interaction  $z_{jj'}=\braOket{\epsilon \ell_{j}m}{z}{4p_{j'}m}$ denotes the matrix element between the final state $j$ and the intermediate state $j'$. The final states $j=1/2$ and $j=3/2$ can be reached from both intermediate states, $j'=1/2$ and $j'=3/2$. In contrast, the final state $j=5/2$ is only reached from the intermediate $j'=3/2$ state and it can not be used to gain temporal information about $\EX$, as illustrated in Fig.~\ref{fig:levelalkali}~(b). 

We use Wigner-Eckart's theorem 
\begin{align}
\braOket{n \ell j m}{z}{n' \ell' j' m} = \nonumber \\
(-1)^{j-m}\threej{j}{1}{j'}{-m}{0}{m}
\braOketred{n \ell j}{\mathbf{r}}{n' \ell' j'}, 
\label{WE}
\end{align}   
and approximate the coupled reduced matrix element, 
\begin{align}
\braOketred{n \ell j}{\mathbf{r}}{n' \ell' j'} 
&\approx \braOket{n\ell}{r}{n'\ell'}\braOketred{j}{\Cone}{j'} \nonumber \\ 
&= \braOketred{n\ell}{\mathbf{r}}{n'\ell'}
\frac{\braOketred{j}{\Cone}{j'}}{\braOketred{\ell}{\Cone}{\ell'}},
\end{align}
using the assumption that the integral over the scalar $r$ is $j$-independent. In this way, we express the matrix element in the $ls$-coupled ($j$) representation in terms of the uncoupled ($ls$) reduced matrix element that can be estimated using non-relativistic many-body perturbation theory. 
The reduced matrix elements of $\mathbf{C^k}$ tensors \cite{lindgren:74:casehfs} for integers is 
\begin{align}
\braOketred{l}{\mathbf{C^k}}{l'} = \nonumber \\(-1)^l[(2l+1)(2l'+1)]^{1/2}\threej{l}{k}{l'}{0}{0}{0}
\end{align}
and for half-integer is 
\begin{align}
\braOketred{j}{\mathbf{C^k}}{j'} = \nonumber \\
(-1)^{j-1/2}[(2j+1)(2j'+1)]^{1/2}\threej{j}{k}{j'}{-1/2}{0}{1/2}.
\end{align}
For simplicity we assume that 
$|\EL(\omega_{j'=3/2\, i})|=|\EL(\omega_{j'=1/2\, i})|$
and that
$|\EX(\omega)|=|\EX(\omega-\wSO)|$. 
The probability density of the final states $|\ck(\epsilon)|^2$ should be summed incoherently to simulate angle-integrated detection of photoelectrons. 
Evaluation of the angular-momentum contribution gives the total probability density for the photoelectrons 
\begin{align}
\sum_{j}|\ck(\epsilon)|^2\approx& \frac{|\EX|^2|\EL|^2}{\hbar^4}\frac{1}{3^4} \braOketred{4p}{\rfat}{4s}|^2 \nonumber \\
\times& 
\Big\{
|\braOketred{\epsilon s}{\rfat}{4p}|^2           \left[ 5+4\cos\Theta \right]  
\nonumber \\
+&
|\braOketred{\epsilon d}{\rfat}{4p}|^2\frac{2}{5}\left[ 8+\cos\Theta \right] 
\Big\},
\label{sumcj}
\end{align}
where the phase of the interference term is given by 
\begin{align}
\Theta(\omega) = & 
\phi_R(\omega_{j'=3/2 \, i})-\phi_R(\omega_{j'=1/2 \, i}) 
\nonumber \\ + & \phiX(\omega-\wSO) - \phiX(\omega)  
\nonumber \\ \approx & \wSO[\tL-\tX(\omega)] ,
\end{align} 
which is proportional to the difference in group delay between $\EL$ and $\EX$ pulses.    
While the laser pulse group delay $\tL$ is determined by the SO excitation energies, the XUV group delay $\tX(\omega)$ is a function of XUV photon energy. In this sense, Eq.~(\ref{sumcj}) shows $\tX(\omega)$  can can be determined by studying photoelectron probability distributions resolved over kinetic energy from coherent SO wave packets.    

In Eq.~(\ref{sumcj}) it is seen that the relative modulation depth of the quantum beat compared to the static background is greater in the $s$-channel than in the $d$-channel. The lesser contrast in the $d$-channel is attributed to the non-modulated $j=5/2$ contribution. In general, however, photoionization to the $d$-channel dominantes over photoionization to the s-channel, due to Fano's propensity rule \cite{FanoPRA1985}, so it is not immediately clear whether the $s$ or $d$ channel have the greatest quantum beat modulations in general, but their modulations will add {\it in phase} as shown by Eq.~(\ref{sumcj}). Eq.~(\ref{sumcj}) is valid for either initial spin polarization, $m=\pm1/2$, which means that an initial statistical mixture will not be a problem for the PANDA method using alkali atoms.  

Excited states of alkali atoms typically have too small SO splitting for quantum beating on the few femtosecond time scale. A more promising approach could be  to use SO {\it valence hole} wave packets in noble gas atoms, generated by tunnel ionization by an ultra-short laser pulse, which have recently been shown to possess good coherence properties \cite{GouliemakisNature2010}.   

%=======================================================================
%

% REFERENCES
%\bibliography{Mendeley.bib}
%\bibliography{MyCollection.bib}

\begin{thebibliography}{58}%
\makeatletter
\providecommand \@ifxundefined [1]{%
 \@ifx{#1\undefined}
}%
\providecommand \@ifnum [1]{%
 \ifnum #1\expandafter \@firstoftwo
 \else \expandafter \@secondoftwo
 \fi
}%
\providecommand \@ifx [1]{%
 \ifx #1\expandafter \@firstoftwo
 \else \expandafter \@secondoftwo
 \fi
}%
\providecommand \natexlab [1]{#1}%
\providecommand \enquote  [1]{``#1''}%
\providecommand \bibnamefont  [1]{#1}%
\providecommand \bibfnamefont [1]{#1}%
\providecommand \citenamefont [1]{#1}%
\providecommand \href@noop [0]{\@secondoftwo}%
\providecommand \href [0]{\begingroup \@sanitize@url \@href}%
\providecommand \@href[1]{\@@startlink{#1}\@@href}%
\providecommand \@@href[1]{\endgroup#1\@@endlink}%
\providecommand \@sanitize@url [0]{\catcode `\\12\catcode `\$12\catcode
  `\&12\catcode `\#12\catcode `\^12\catcode `\_12\catcode `\%12\relax}%
\providecommand \@@startlink[1]{}%
\providecommand \@@endlink[0]{}%
\providecommand \url  [0]{\begingroup\@sanitize@url \@url }%
\providecommand \@url [1]{\endgroup\@href {#1}{\urlprefix }}%
\providecommand \urlprefix  [0]{URL }%
\providecommand \Eprint [0]{\href }%
\providecommand \doibase [0]{http://dx.doi.org/}%
\providecommand \selectlanguage [0]{\@gobble}%
\providecommand \bibinfo  [0]{\@secondoftwo}%
\providecommand \bibfield  [0]{\@secondoftwo}%
\providecommand \translation [1]{[#1]}%
\providecommand \BibitemOpen [0]{}%
\providecommand \bibitemStop [0]{}%
\providecommand \bibitemNoStop [0]{.\EOS\space}%
\providecommand \EOS [0]{\spacefactor3000\relax}%
\providecommand \BibitemShut  [1]{\csname bibitem#1\endcsname}%
\let\auto@bib@innerbib\@empty
%</preamble>
\bibitem [{\citenamefont {Zewail}(2000)}]{Zewail2000}%
  \BibitemOpen
  \bibfield  {author} {\bibinfo {author} {\bibfnamefont {Ahmed~H.}\
  \bibnamefont {Zewail}},\ }\bibfield  {title} {\enquote {\bibinfo {title}
  {{Femtochemistry: Atomic-Scale Dynamics of the Chemical Bond}},}\ }\href
  {\doibase 10.1021/jp001460h} {\bibfield  {journal} {\bibinfo  {journal} {The
  Journal of Physical Chemistry A}\ }\textbf {\bibinfo {volume} {104}},\
  \bibinfo {pages} {5660--5694} (\bibinfo {year} {2000})}\BibitemShut {NoStop}%
\bibitem [{\citenamefont {Strickland}\ and\ \citenamefont
  {Mourou}(1985)}]{Strickland1985}%
  \BibitemOpen
  \bibfield  {author} {\bibinfo {author} {\bibfnamefont {Donna}\ \bibnamefont
  {Strickland}}\ and\ \bibinfo {author} {\bibfnamefont {Gerard}\ \bibnamefont
  {Mourou}},\ }\bibfield  {title} {\enquote {\bibinfo {title} {{Compression of
  amplified chirped optical pulses}},}\ }\href {\doibase
  10.1016/0030-4018(85)90120-8} {\bibfield  {journal} {\bibinfo  {journal}
  {Optics Communications}\ }\textbf {\bibinfo {volume} {56}},\ \bibinfo {pages}
  {219--221} (\bibinfo {year} {1985})},\ \Eprint
  {http://arxiv.org/abs/arXiv:1011.1669v3} {arXiv:arXiv:1011.1669v3}
  \BibitemShut {NoStop}%
\bibitem [{\citenamefont {Lewenstein}\ \emph {et~al.}(1994)\citenamefont
  {Lewenstein}, \citenamefont {Balcou}, \citenamefont {Ivanov}, \citenamefont
  {L'Huillier},\ and\ \citenamefont {Corkum}}]{LewensteinPRA1994}%
  \BibitemOpen
  \bibfield  {author} {\bibinfo {author} {\bibfnamefont {M}~\bibnamefont
  {Lewenstein}}, \bibinfo {author} {\bibfnamefont {Ph.}\ \bibnamefont
  {Balcou}}, \bibinfo {author} {\bibfnamefont {M~Yu.}\ \bibnamefont {Ivanov}},
  \bibinfo {author} {\bibfnamefont {Anne}\ \bibnamefont {L'Huillier}}, \ and\
  \bibinfo {author} {\bibfnamefont {P~B}\ \bibnamefont {Corkum}},\ }\bibfield
  {title} {\enquote {\bibinfo {title} {{Theory of high-harmonic generation by
  low-frequency laser fields}},}\ }\href {\doibase 10.1103/PhysRevA.49.2117}
  {\bibfield  {journal} {\bibinfo  {journal} {Phys. Rev. A}\ }\textbf {\bibinfo
  {volume} {49}},\ \bibinfo {pages} {2117--2132} (\bibinfo {year}
  {1994})}\BibitemShut {NoStop}%
\bibitem [{\citenamefont {Chen}\ \emph {et~al.}(2010)\citenamefont {Chen},
  \citenamefont {Arpin}, \citenamefont {Popmintchev}, \citenamefont {Gerrity},
  \citenamefont {Zhang}, \citenamefont {Seaberg}, \citenamefont {Popmintchev},
  \citenamefont {Murnane},\ and\ \citenamefont {Kapteyn}}]{Chen2010}%
  \BibitemOpen
  \bibfield  {author} {\bibinfo {author} {\bibfnamefont {M.~C.}\ \bibnamefont
  {Chen}}, \bibinfo {author} {\bibfnamefont {P.}~\bibnamefont {Arpin}},
  \bibinfo {author} {\bibfnamefont {T.}~\bibnamefont {Popmintchev}}, \bibinfo
  {author} {\bibfnamefont {M.}~\bibnamefont {Gerrity}}, \bibinfo {author}
  {\bibfnamefont {B.}~\bibnamefont {Zhang}}, \bibinfo {author} {\bibfnamefont
  {M.}~\bibnamefont {Seaberg}}, \bibinfo {author} {\bibfnamefont
  {D.}~\bibnamefont {Popmintchev}}, \bibinfo {author} {\bibfnamefont {M.~M.}\
  \bibnamefont {Murnane}}, \ and\ \bibinfo {author} {\bibfnamefont {H.~C.}\
  \bibnamefont {Kapteyn}},\ }\bibfield  {title} {\enquote {\bibinfo {title}
  {{Bright, coherent, ultrafast soft x-ray harmonics spanning the water window
  from a tabletop light source}},}\ }\href {\doibase
  10.1103/PhysRevLett.105.173901} {\bibfield  {journal} {\bibinfo  {journal}
  {Physical Review Letters}\ }\textbf {\bibinfo {volume} {105}},\ \bibinfo
  {pages} {173901} (\bibinfo {year} {2010})},\ \Eprint
  {http://arxiv.org/abs/1006.3942} {arXiv:1006.3942} \BibitemShut {NoStop}%
\bibitem [{\citenamefont {Paul}\ \emph {et~al.}(2001)\citenamefont {Paul},
  \citenamefont {Toma}, \citenamefont {Breger}, \citenamefont {Mullot},
  \citenamefont {Aug{\'{e}}}, \citenamefont {Balcou}, \citenamefont {Muller},\
  and\ \citenamefont {Agostini}}]{PaulScience2001}%
  \BibitemOpen
  \bibfield  {author} {\bibinfo {author} {\bibfnamefont {P~M}\ \bibnamefont
  {Paul}}, \bibinfo {author} {\bibfnamefont {E~S}\ \bibnamefont {Toma}},
  \bibinfo {author} {\bibfnamefont {P}~\bibnamefont {Breger}}, \bibinfo
  {author} {\bibfnamefont {G}~\bibnamefont {Mullot}}, \bibinfo {author}
  {\bibfnamefont {F}~\bibnamefont {Aug{\'{e}}}}, \bibinfo {author}
  {\bibfnamefont {Ph.}\ \bibnamefont {Balcou}}, \bibinfo {author}
  {\bibfnamefont {H~G}\ \bibnamefont {Muller}}, \ and\ \bibinfo {author}
  {\bibfnamefont {P}~\bibnamefont {Agostini}},\ }\bibfield  {title} {\enquote
  {\bibinfo {title} {{Observation of a Train of Attosecond Pulses from High
  Harmonic Generation}},}\ }\href {\doibase 10.1126/science.1059413} {\bibfield
   {journal} {\bibinfo  {journal} {Science}\ }\textbf {\bibinfo {volume}
  {292}},\ \bibinfo {pages} {1689--1692} (\bibinfo {year} {2001})}\BibitemShut
  {NoStop}%
\bibitem [{\citenamefont {Hentschel}\ \emph {et~al.}(2001)\citenamefont
  {Hentschel}, \citenamefont {Kienberger}, \citenamefont {Spielmann},
  \citenamefont {Reider}, \citenamefont {Milosevic}, \citenamefont {Brabec},
  \citenamefont {Corkum}, \citenamefont {Heinzmann}, \citenamefont {Drescher},\
  and\ \citenamefont {Krausz}}]{HentschelNature2001}%
  \BibitemOpen
  \bibfield  {author} {\bibinfo {author} {\bibfnamefont {M}~\bibnamefont
  {Hentschel}}, \bibinfo {author} {\bibfnamefont {R}~\bibnamefont
  {Kienberger}}, \bibinfo {author} {\bibfnamefont {Ch.}\ \bibnamefont
  {Spielmann}}, \bibinfo {author} {\bibfnamefont {G~A}\ \bibnamefont {Reider}},
  \bibinfo {author} {\bibfnamefont {N}~\bibnamefont {Milosevic}}, \bibinfo
  {author} {\bibfnamefont {T}~\bibnamefont {Brabec}}, \bibinfo {author}
  {\bibfnamefont {P}~\bibnamefont {Corkum}}, \bibinfo {author} {\bibfnamefont
  {U}~\bibnamefont {Heinzmann}}, \bibinfo {author} {\bibfnamefont
  {M}~\bibnamefont {Drescher}}, \ and\ \bibinfo {author} {\bibfnamefont
  {F}~\bibnamefont {Krausz}},\ }\bibfield  {title} {\enquote {\bibinfo {title}
  {{Attosecond metrology}},}\ }\href@noop {} {\bibfield  {journal} {\bibinfo
  {journal} {Nature}\ }\textbf {\bibinfo {volume} {414}},\ \bibinfo {pages}
  {509--513} (\bibinfo {year} {2001})}\BibitemShut {NoStop}%
\bibitem [{\citenamefont {Krausz}\ and\ \citenamefont
  {Ivanov}(2009)}]{Krausz2009}%
  \BibitemOpen
  \bibfield  {author} {\bibinfo {author} {\bibfnamefont {Ferenc}\ \bibnamefont
  {Krausz}}\ and\ \bibinfo {author} {\bibfnamefont {Misha}\ \bibnamefont
  {Ivanov}},\ }\bibfield  {title} {\enquote {\bibinfo {title} {{Attosecond
  physics}},}\ }\href {\doibase 10.1103/RevModPhys.81.163} {\bibfield
  {journal} {\bibinfo  {journal} {Reviews of Modern Physics}\ }\textbf
  {\bibinfo {volume} {81}},\ \bibinfo {pages} {163--234} (\bibinfo {year}
  {2009})}\BibitemShut {NoStop}%
\bibitem [{\citenamefont {Wirth}\ \emph {et~al.}(2011)\citenamefont {Wirth},
  \citenamefont {Hassan}, \citenamefont {Grguras}, \citenamefont {Gagnon},
  \citenamefont {Moulet}, \citenamefont {Luu}, \citenamefont {Pabst},
  \citenamefont {Santra}, \citenamefont {Alahmed}, \citenamefont {Azzeer},
  \citenamefont {Yakovlev}, \citenamefont {Pervak}, \citenamefont {Krausz},\
  and\ \citenamefont {Goulielmakis}}]{Wirth2011}%
  \BibitemOpen
  \bibfield  {author} {\bibinfo {author} {\bibfnamefont {A.}~\bibnamefont
  {Wirth}}, \bibinfo {author} {\bibfnamefont {M.~T.}\ \bibnamefont {Hassan}},
  \bibinfo {author} {\bibfnamefont {I.}~\bibnamefont {Grguras}}, \bibinfo
  {author} {\bibfnamefont {J.}~\bibnamefont {Gagnon}}, \bibinfo {author}
  {\bibfnamefont {A.}~\bibnamefont {Moulet}}, \bibinfo {author} {\bibfnamefont
  {T.~T.}\ \bibnamefont {Luu}}, \bibinfo {author} {\bibfnamefont
  {S.}~\bibnamefont {Pabst}}, \bibinfo {author} {\bibfnamefont
  {R.}~\bibnamefont {Santra}}, \bibinfo {author} {\bibfnamefont {Z.~A.}\
  \bibnamefont {Alahmed}}, \bibinfo {author} {\bibfnamefont {A.~M.}\
  \bibnamefont {Azzeer}}, \bibinfo {author} {\bibfnamefont {V.~S.}\
  \bibnamefont {Yakovlev}}, \bibinfo {author} {\bibfnamefont {V.}~\bibnamefont
  {Pervak}}, \bibinfo {author} {\bibfnamefont {F.}~\bibnamefont {Krausz}}, \
  and\ \bibinfo {author} {\bibfnamefont {E.}~\bibnamefont {Goulielmakis}},\
  }\bibfield  {title} {\enquote {\bibinfo {title} {{Synthesized Light
  Transients}},}\ }\href {\doibase 10.1126/science.1210268} {\bibfield
  {journal} {\bibinfo  {journal} {Science}\ }\textbf {\bibinfo {volume}
  {334}},\ \bibinfo {pages} {195--200} (\bibinfo {year} {2011})}\BibitemShut
  {NoStop}%
\bibitem [{\citenamefont {Mairesse}\ and\ \citenamefont
  {Qu{\'{e}}r{\'{e}}}(2005)}]{MairessePRA2005}%
  \BibitemOpen
  \bibfield  {author} {\bibinfo {author} {\bibfnamefont {Y}~\bibnamefont
  {Mairesse}}\ and\ \bibinfo {author} {\bibfnamefont {F}~\bibnamefont
  {Qu{\'{e}}r{\'{e}}}},\ }\bibfield  {title} {\enquote {\bibinfo {title}
  {{Frequency-resolved optical gating for complete reconstruction of attosecond
  bursts}},}\ }\href {\doibase 10.1103/PhysRevA.71.011401} {\bibfield
  {journal} {\bibinfo  {journal} {Phys. Rev. A}\ }\textbf {\bibinfo {volume}
  {71}},\ \bibinfo {pages} {11401} (\bibinfo {year} {2005})}\BibitemShut
  {NoStop}%
\bibitem [{\citenamefont {Goulielmakis}\ \emph {et~al.}(2010)\citenamefont
  {Goulielmakis}, \citenamefont {Loh}, \citenamefont {Wirth}, \citenamefont
  {Santra}, \citenamefont {Rohringer}, \citenamefont {Yakovlev}, \citenamefont
  {Zherebtsov}, \citenamefont {Pfeifer}, \citenamefont {Azzeer}, \citenamefont
  {Kling}, \citenamefont {Leone},\ and\ \citenamefont
  {Krausz}}]{GouliemakisNature2010}%
  \BibitemOpen
  \bibfield  {author} {\bibinfo {author} {\bibfnamefont {Eleftherios}\
  \bibnamefont {Goulielmakis}}, \bibinfo {author} {\bibfnamefont {Zhi-Heng}\
  \bibnamefont {Loh}}, \bibinfo {author} {\bibfnamefont {Adrian}\ \bibnamefont
  {Wirth}}, \bibinfo {author} {\bibfnamefont {Robin}\ \bibnamefont {Santra}},
  \bibinfo {author} {\bibfnamefont {Nina}\ \bibnamefont {Rohringer}}, \bibinfo
  {author} {\bibfnamefont {Vladislav~S}\ \bibnamefont {Yakovlev}}, \bibinfo
  {author} {\bibfnamefont {Sergey}\ \bibnamefont {Zherebtsov}}, \bibinfo
  {author} {\bibfnamefont {Thomas}\ \bibnamefont {Pfeifer}}, \bibinfo {author}
  {\bibfnamefont {Abdallah~M}\ \bibnamefont {Azzeer}}, \bibinfo {author}
  {\bibfnamefont {Matthias~F}\ \bibnamefont {Kling}}, \bibinfo {author}
  {\bibfnamefont {Stephen~R}\ \bibnamefont {Leone}}, \ and\ \bibinfo {author}
  {\bibfnamefont {Ferenc}\ \bibnamefont {Krausz}},\ }\bibfield  {title}
  {\enquote {\bibinfo {title} {{Real-time observation of valence electron
  motion}},}\ }\href@noop {} {\bibfield  {journal} {\bibinfo  {journal}
  {Nature}\ }\textbf {\bibinfo {volume} {466}},\ \bibinfo {pages} {739--743}
  (\bibinfo {year} {2010})}\BibitemShut {NoStop}%
\bibitem [{\citenamefont {Worner}\ \emph {et~al.}(2010)\citenamefont {Worner},
  \citenamefont {Bertrand}, \citenamefont {Kartashov}, \citenamefont {Corkum},\
  and\ \citenamefont {Villeneuve}}]{WornerNature2010}%
  \BibitemOpen
  \bibfield  {author} {\bibinfo {author} {\bibfnamefont {H~J}\ \bibnamefont
  {Worner}}, \bibinfo {author} {\bibfnamefont {J~B}\ \bibnamefont {Bertrand}},
  \bibinfo {author} {\bibfnamefont {D~V}\ \bibnamefont {Kartashov}}, \bibinfo
  {author} {\bibfnamefont {P~B}\ \bibnamefont {Corkum}}, \ and\ \bibinfo
  {author} {\bibfnamefont {D~M}\ \bibnamefont {Villeneuve}},\ }\bibfield
  {title} {\enquote {\bibinfo {title} {{Following a chemical reaction using
  high-harmonic interferometry}},}\ }\href@noop {} {\bibfield  {journal}
  {\bibinfo  {journal} {Nature}\ }\textbf {\bibinfo {volume} {466}},\ \bibinfo
  {pages} {604--607} (\bibinfo {year} {2010})}\BibitemShut {NoStop}%
\bibitem [{\citenamefont {Schultze}\ \emph {et~al.}(2010)\citenamefont
  {Schultze}, \citenamefont {Fiess}, \citenamefont {Karpowicz}, \citenamefont
  {Gagnon}, \citenamefont {Korbman}, \citenamefont {Hofstetter}, \citenamefont
  {Neppl}, \citenamefont {Cavalieri}, \citenamefont {Komninos}, \citenamefont
  {Mercouris}, \citenamefont {Nicolaides}, \citenamefont {Pazourek},
  \citenamefont {Nagele}, \citenamefont {Feist}, \citenamefont
  {Burgd{\"{o}}rfer}, \citenamefont {Azzeer}, \citenamefont {Ernstorfer},
  \citenamefont {Kienberger}, \citenamefont {Kleineberg}, \citenamefont
  {Goulielmakis}, \citenamefont {Krausz},\ and\ \citenamefont
  {Yakovlev}}]{SchultzeScience2010}%
  \BibitemOpen
  \bibfield  {author} {\bibinfo {author} {\bibfnamefont {M}~\bibnamefont
  {Schultze}}, \bibinfo {author} {\bibfnamefont {M}~\bibnamefont {Fiess}},
  \bibinfo {author} {\bibfnamefont {N}~\bibnamefont {Karpowicz}}, \bibinfo
  {author} {\bibfnamefont {J}~\bibnamefont {Gagnon}}, \bibinfo {author}
  {\bibfnamefont {M}~\bibnamefont {Korbman}}, \bibinfo {author} {\bibfnamefont
  {M}~\bibnamefont {Hofstetter}}, \bibinfo {author} {\bibfnamefont
  {S}~\bibnamefont {Neppl}}, \bibinfo {author} {\bibfnamefont {A~L}\
  \bibnamefont {Cavalieri}}, \bibinfo {author} {\bibfnamefont {Y}~\bibnamefont
  {Komninos}}, \bibinfo {author} {\bibfnamefont {Th.}\ \bibnamefont
  {Mercouris}}, \bibinfo {author} {\bibfnamefont {C~A}\ \bibnamefont
  {Nicolaides}}, \bibinfo {author} {\bibfnamefont {R}~\bibnamefont {Pazourek}},
  \bibinfo {author} {\bibfnamefont {S}~\bibnamefont {Nagele}}, \bibinfo
  {author} {\bibfnamefont {J}~\bibnamefont {Feist}}, \bibinfo {author}
  {\bibfnamefont {J}~\bibnamefont {Burgd{\"{o}}rfer}}, \bibinfo {author}
  {\bibfnamefont {A~M}\ \bibnamefont {Azzeer}}, \bibinfo {author}
  {\bibfnamefont {R}~\bibnamefont {Ernstorfer}}, \bibinfo {author}
  {\bibfnamefont {R}~\bibnamefont {Kienberger}}, \bibinfo {author}
  {\bibfnamefont {U}~\bibnamefont {Kleineberg}}, \bibinfo {author}
  {\bibfnamefont {E}~\bibnamefont {Goulielmakis}}, \bibinfo {author}
  {\bibfnamefont {F}~\bibnamefont {Krausz}}, \ and\ \bibinfo {author}
  {\bibfnamefont {V~S}\ \bibnamefont {Yakovlev}},\ }\bibfield  {title}
  {\enquote {\bibinfo {title} {{Delay in Photoemission}},}\ }\href@noop {}
  {\bibfield  {journal} {\bibinfo  {journal} {Science}\ }\textbf {\bibinfo
  {volume} {328}},\ \bibinfo {pages} {1658--1662} (\bibinfo {year}
  {2010})}\BibitemShut {NoStop}%
\bibitem [{\citenamefont {Kl{\"{u}}nder}\ \emph {et~al.}(2011)\citenamefont
  {Kl{\"{u}}nder}, \citenamefont {Dahlstr{\"{o}}m}, \citenamefont
  {Gisselbrecht}, \citenamefont {Fordell}, \citenamefont {Swoboda},
  \citenamefont {Gu{\'{e}}not}, \citenamefont {Johnsson}, \citenamefont
  {Caillat}, \citenamefont {Mauritsson}, \citenamefont {Maquet}, \citenamefont
  {Ta{\"{i}}eb},\ and\ \citenamefont {L'Huillier}}]{KlunderPRL2011}%
  \BibitemOpen
  \bibfield  {author} {\bibinfo {author} {\bibfnamefont {K}~\bibnamefont
  {Kl{\"{u}}nder}}, \bibinfo {author} {\bibfnamefont {J~M}\ \bibnamefont
  {Dahlstr{\"{o}}m}}, \bibinfo {author} {\bibfnamefont {M}~\bibnamefont
  {Gisselbrecht}}, \bibinfo {author} {\bibfnamefont {T}~\bibnamefont
  {Fordell}}, \bibinfo {author} {\bibfnamefont {M}~\bibnamefont {Swoboda}},
  \bibinfo {author} {\bibfnamefont {D}~\bibnamefont {Gu{\'{e}}not}}, \bibinfo
  {author} {\bibfnamefont {P}~\bibnamefont {Johnsson}}, \bibinfo {author}
  {\bibfnamefont {J}~\bibnamefont {Caillat}}, \bibinfo {author} {\bibfnamefont
  {J}~\bibnamefont {Mauritsson}}, \bibinfo {author} {\bibfnamefont
  {A}~\bibnamefont {Maquet}}, \bibinfo {author} {\bibfnamefont {R}~\bibnamefont
  {Ta{\"{i}}eb}}, \ and\ \bibinfo {author} {\bibfnamefont {A}~\bibnamefont
  {L'Huillier}},\ }\bibfield  {title} {\enquote {\bibinfo {title} {{Probing
  Single-Photon Ionization on the Attosecond Time Scale}},}\ }\href {\doibase
  10.1103/PhysRevLett.106.143002} {\bibfield  {journal} {\bibinfo  {journal}
  {Phys. Rev. Lett.}\ }\textbf {\bibinfo {volume} {106}},\ \bibinfo {pages}
  {143002} (\bibinfo {year} {2011})}\BibitemShut {NoStop}%
\bibitem [{\citenamefont {Ossiander}\ \emph {et~al.}(2016)\citenamefont
  {Ossiander}, \citenamefont {Siegrist}, \citenamefont {Shirvanyan},
  \citenamefont {Pazourek}, \citenamefont {Sommer}, \citenamefont {Latka},
  \citenamefont {Guggenmos}, \citenamefont {Nagele}, \citenamefont {Feist},
  \citenamefont {Burgd{\"{o}}rfer}, \citenamefont {Kienberger},\ and\
  \citenamefont {Schultze}}]{Ossiander2016}%
  \BibitemOpen
  \bibfield  {author} {\bibinfo {author} {\bibfnamefont {M.}~\bibnamefont
  {Ossiander}}, \bibinfo {author} {\bibfnamefont {F.}~\bibnamefont {Siegrist}},
  \bibinfo {author} {\bibfnamefont {V.}~\bibnamefont {Shirvanyan}}, \bibinfo
  {author} {\bibfnamefont {R.}~\bibnamefont {Pazourek}}, \bibinfo {author}
  {\bibfnamefont {A.}~\bibnamefont {Sommer}}, \bibinfo {author} {\bibfnamefont
  {T.}~\bibnamefont {Latka}}, \bibinfo {author} {\bibfnamefont
  {A.}~\bibnamefont {Guggenmos}}, \bibinfo {author} {\bibfnamefont
  {S.}~\bibnamefont {Nagele}}, \bibinfo {author} {\bibfnamefont
  {J.}~\bibnamefont {Feist}}, \bibinfo {author} {\bibfnamefont
  {J.}~\bibnamefont {Burgd{\"{o}}rfer}}, \bibinfo {author} {\bibfnamefont
  {R.}~\bibnamefont {Kienberger}}, \ and\ \bibinfo {author} {\bibfnamefont
  {M.}~\bibnamefont {Schultze}},\ }\bibfield  {title} {\enquote {\bibinfo
  {title} {{Attosecond correlation dynamics}},}\ }\href {\doibase
  10.1038/nphys3941} {\bibfield  {journal} {\bibinfo  {journal} {Nature
  Physics}\ }\textbf {\bibinfo {volume} {13}},\ \bibinfo {pages} {280--285}
  (\bibinfo {year} {2016})},\ \Eprint {http://arxiv.org/abs/0803.0582}
  {arXiv:0803.0582} \BibitemShut {NoStop}%
\bibitem [{\citenamefont {Isinger}\ \emph {et~al.}(2017)\citenamefont
  {Isinger}, \citenamefont {Squibb}, \citenamefont {Busto}, \citenamefont
  {Zhong}, \citenamefont {Harth}, \citenamefont {Kroon}, \citenamefont {Nandi},
  \citenamefont {Arnold}, \citenamefont {Miranda}, \citenamefont
  {Dahlstr{\"{o}}m}, \citenamefont {Lindroth}, \citenamefont {Feifel},
  \citenamefont {Gisselbrecht},\ and\ \citenamefont
  {L'Huillier}}]{Isinger2017}%
  \BibitemOpen
  \bibfield  {author} {\bibinfo {author} {\bibfnamefont {M.}~\bibnamefont
  {Isinger}}, \bibinfo {author} {\bibfnamefont {R.~J.}\ \bibnamefont {Squibb}},
  \bibinfo {author} {\bibfnamefont {D.}~\bibnamefont {Busto}}, \bibinfo
  {author} {\bibfnamefont {S.}~\bibnamefont {Zhong}}, \bibinfo {author}
  {\bibfnamefont {A.}~\bibnamefont {Harth}}, \bibinfo {author} {\bibfnamefont
  {D.}~\bibnamefont {Kroon}}, \bibinfo {author} {\bibfnamefont
  {S.}~\bibnamefont {Nandi}}, \bibinfo {author} {\bibfnamefont {C.~L.}\
  \bibnamefont {Arnold}}, \bibinfo {author} {\bibfnamefont {M.}~\bibnamefont
  {Miranda}}, \bibinfo {author} {\bibfnamefont {J.~M.}\ \bibnamefont
  {Dahlstr{\"{o}}m}}, \bibinfo {author} {\bibfnamefont {E.}~\bibnamefont
  {Lindroth}}, \bibinfo {author} {\bibfnamefont {R.}~\bibnamefont {Feifel}},
  \bibinfo {author} {\bibfnamefont {M.}~\bibnamefont {Gisselbrecht}}, \ and\
  \bibinfo {author} {\bibfnamefont {A.}~\bibnamefont {L'Huillier}},\ }\bibfield
   {title} {\enquote {\bibinfo {title} {{Photoionization in the time and
  frequency domain}},}\ }\href {\doibase 10.1126/science.aao7043} {\bibfield
  {journal} {\bibinfo  {journal} {Science}\ }\textbf {\bibinfo {volume}
  {358}},\ \bibinfo {pages} {893--896} (\bibinfo {year} {2017})},\ \Eprint
  {http://arxiv.org/abs/1709.01780} {arXiv:1709.01780} \BibitemShut {NoStop}%
\bibitem [{\citenamefont {Huppert}\ \emph {et~al.}(2016)\citenamefont
  {Huppert}, \citenamefont {Jordan}, \citenamefont {Baykusheva}, \citenamefont
  {{Von Conta}},\ and\ \citenamefont {W{\"{o}}rner}}]{Huppert2016}%
  \BibitemOpen
  \bibfield  {author} {\bibinfo {author} {\bibfnamefont {Martin}\ \bibnamefont
  {Huppert}}, \bibinfo {author} {\bibfnamefont {Inga}\ \bibnamefont {Jordan}},
  \bibinfo {author} {\bibfnamefont {Denitsa}\ \bibnamefont {Baykusheva}},
  \bibinfo {author} {\bibfnamefont {Aaron}\ \bibnamefont {{Von Conta}}}, \ and\
  \bibinfo {author} {\bibfnamefont {Hans~Jakob}\ \bibnamefont {W{\"{o}}rner}},\
  }\bibfield  {title} {\enquote {\bibinfo {title} {{Attosecond Delays in
  Molecular Photoionization}},}\ }\href {\doibase
  10.1103/PhysRevLett.117.093001} {\bibfield  {journal} {\bibinfo  {journal}
  {Physical Review Letters}\ }\textbf {\bibinfo {volume} {117}},\ \bibinfo
  {pages} {093001} (\bibinfo {year} {2016})},\ \Eprint
  {http://arxiv.org/abs/1607.07435} {arXiv:1607.07435} \BibitemShut {NoStop}%
\bibitem [{\citenamefont {Cavalieri}\ \emph {et~al.}(2007)\citenamefont
  {Cavalieri}, \citenamefont {M{\"{u}}ller}, \citenamefont {Uphues},
  \citenamefont {Yakovlev}, \citenamefont {Baltu{\v{s}}ka}, \citenamefont
  {Horvath}, \citenamefont {Schmidt}, \citenamefont {Bl{\"{u}}mel},
  \citenamefont {Holzwarth}, \citenamefont {Hendel}, \citenamefont {Drescher},
  \citenamefont {Kleineberg}, \citenamefont {Echenique}, \citenamefont
  {Kienberger}, \citenamefont {Krausz},\ and\ \citenamefont
  {Heinzmann}}]{Cavalieri2007}%
  \BibitemOpen
  \bibfield  {author} {\bibinfo {author} {\bibfnamefont {A.~L.}\ \bibnamefont
  {Cavalieri}}, \bibinfo {author} {\bibfnamefont {N.}~\bibnamefont
  {M{\"{u}}ller}}, \bibinfo {author} {\bibfnamefont {Th.}\ \bibnamefont
  {Uphues}}, \bibinfo {author} {\bibfnamefont {V.~S.}\ \bibnamefont
  {Yakovlev}}, \bibinfo {author} {\bibfnamefont {A.}~\bibnamefont
  {Baltu{\v{s}}ka}}, \bibinfo {author} {\bibfnamefont {B.}~\bibnamefont
  {Horvath}}, \bibinfo {author} {\bibfnamefont {B.}~\bibnamefont {Schmidt}},
  \bibinfo {author} {\bibfnamefont {L.}~\bibnamefont {Bl{\"{u}}mel}}, \bibinfo
  {author} {\bibfnamefont {R.}~\bibnamefont {Holzwarth}}, \bibinfo {author}
  {\bibfnamefont {S.}~\bibnamefont {Hendel}}, \bibinfo {author} {\bibfnamefont
  {M.}~\bibnamefont {Drescher}}, \bibinfo {author} {\bibfnamefont
  {U.}~\bibnamefont {Kleineberg}}, \bibinfo {author} {\bibfnamefont {P.~M.}\
  \bibnamefont {Echenique}}, \bibinfo {author} {\bibfnamefont {R.}~\bibnamefont
  {Kienberger}}, \bibinfo {author} {\bibfnamefont {F.}~\bibnamefont {Krausz}},
  \ and\ \bibinfo {author} {\bibfnamefont {U.}~\bibnamefont {Heinzmann}},\
  }\bibfield  {title} {\enquote {\bibinfo {title} {{Attosecond spectroscopy in
  condensed matter}},}\ }\href {\doibase 10.1038/nature06229} {\bibfield
  {journal} {\bibinfo  {journal} {Nature}\ }\textbf {\bibinfo {volume} {449}},\
  \bibinfo {pages} {1029--1032} (\bibinfo {year} {2007})}\BibitemShut {NoStop}%
\bibitem [{\citenamefont {Neppl}\ \emph {et~al.}(2015)\citenamefont {Neppl},
  \citenamefont {Ernstorfer}, \citenamefont {Cavalieri}, \citenamefont
  {Lemell}, \citenamefont {Wachter}, \citenamefont {Magerl}, \citenamefont
  {Bothschafter}, \citenamefont {Jobst}, \citenamefont {Hofstetter},
  \citenamefont {Kleineberg}, \citenamefont {Barth}, \citenamefont {Menzel},
  \citenamefont {Burgd{\"{o}}rfer}, \citenamefont {Feulner}, \citenamefont
  {Krausz},\ and\ \citenamefont {Kienberger}}]{Neppl2015}%
  \BibitemOpen
  \bibfield  {author} {\bibinfo {author} {\bibfnamefont {S.}~\bibnamefont
  {Neppl}}, \bibinfo {author} {\bibfnamefont {R.}~\bibnamefont {Ernstorfer}},
  \bibinfo {author} {\bibfnamefont {A.~L.}\ \bibnamefont {Cavalieri}}, \bibinfo
  {author} {\bibfnamefont {C.}~\bibnamefont {Lemell}}, \bibinfo {author}
  {\bibfnamefont {G.}~\bibnamefont {Wachter}}, \bibinfo {author} {\bibfnamefont
  {E.}~\bibnamefont {Magerl}}, \bibinfo {author} {\bibfnamefont {E.~M.}\
  \bibnamefont {Bothschafter}}, \bibinfo {author} {\bibfnamefont
  {M.}~\bibnamefont {Jobst}}, \bibinfo {author} {\bibfnamefont
  {M.}~\bibnamefont {Hofstetter}}, \bibinfo {author} {\bibfnamefont
  {U.}~\bibnamefont {Kleineberg}}, \bibinfo {author} {\bibfnamefont {J.~V.}\
  \bibnamefont {Barth}}, \bibinfo {author} {\bibfnamefont {D.}~\bibnamefont
  {Menzel}}, \bibinfo {author} {\bibfnamefont {J.}~\bibnamefont
  {Burgd{\"{o}}rfer}}, \bibinfo {author} {\bibfnamefont {P.}~\bibnamefont
  {Feulner}}, \bibinfo {author} {\bibfnamefont {F.}~\bibnamefont {Krausz}}, \
  and\ \bibinfo {author} {\bibfnamefont {R.}~\bibnamefont {Kienberger}},\
  }\bibfield  {title} {\enquote {\bibinfo {title} {{Direct observation of
  electron propagation and dielectric screening on the atomic length scale}},}\
  }\href {\doibase 10.1038/nature14094} {\bibfield  {journal} {\bibinfo
  {journal} {Nature}\ }\textbf {\bibinfo {volume} {517}},\ \bibinfo {pages}
  {342--346} (\bibinfo {year} {2015})},\ \Eprint
  {http://arxiv.org/abs/NIHMS150003} {arXiv:NIHMS150003} \BibitemShut {NoStop}%
\bibitem [{\citenamefont {Dahlstr{\"{o}}m}\ \emph {et~al.}(2011)\citenamefont
  {Dahlstr{\"{o}}m}, \citenamefont {L'Huillier},\ and\ \citenamefont
  {Mauritsson}}]{DahlstromJPB2011}%
  \BibitemOpen
  \bibfield  {author} {\bibinfo {author} {\bibfnamefont {J~M}\ \bibnamefont
  {Dahlstr{\"{o}}m}}, \bibinfo {author} {\bibfnamefont {A}~\bibnamefont
  {L'Huillier}}, \ and\ \bibinfo {author} {\bibfnamefont {J}~\bibnamefont
  {Mauritsson}},\ }\bibfield  {title} {\enquote {\bibinfo {title} {{Quantum
  mechanical approach to probing the birth of attosecond pulses using a
  two-colour field}},}\ }\href
  {http://stacks.iop.org/0953-4075/44/i=9/a=095602} {\bibfield  {journal}
  {\bibinfo  {journal} {Journal of Physics B: Atomic, Molecular and Optical
  Physics}\ }\textbf {\bibinfo {volume} {44}},\ \bibinfo {pages} {95602}
  (\bibinfo {year} {2011})}\BibitemShut {NoStop}%
\bibitem [{\citenamefont {Pabst}\ and\ \citenamefont
  {Dahlstr{\"{o}}m}(2016)}]{PabstPRA2016}%
  \BibitemOpen
  \bibfield  {author} {\bibinfo {author} {\bibfnamefont {Stefan}\ \bibnamefont
  {Pabst}}\ and\ \bibinfo {author} {\bibfnamefont {Jan~Marcus}\ \bibnamefont
  {Dahlstr{\"{o}}m}},\ }\bibfield  {title} {\enquote {\bibinfo {title}
  {{Eliminating the dipole phase in attosecond pulse characterization using
  Rydberg wave packets}},}\ }\href {\doibase 10.1103/PhysRevA.94.013411}
  {\bibfield  {journal} {\bibinfo  {journal} {Physical Review A - Atomic,
  Molecular, and Optical Physics}\ }\textbf {\bibinfo {volume} {94}},\ \bibinfo
  {pages} {13411} (\bibinfo {year} {2016})}\BibitemShut {NoStop}%
\bibitem [{\citenamefont {Trebino}(2002)}]{FROG}%
  \BibitemOpen
  \bibfield  {author} {\bibinfo {author} {\bibfnamefont {Rick}\ \bibnamefont
  {Trebino}},\ }\href@noop {} {\emph {\bibinfo {title} {{Frequency-Resolved
  Optical Gating: The Measurement of Ultrashort Laser Pulses}}}}\ (\bibinfo
  {publisher} {Springer US},\ \bibinfo {year} {2002})\BibitemShut {NoStop}%
\bibitem [{\citenamefont {Iaconis}\ and\ \citenamefont
  {Walmsley}(1998)}]{Iaconis1998}%
  \BibitemOpen
  \bibfield  {author} {\bibinfo {author} {\bibfnamefont {C.}~\bibnamefont
  {Iaconis}}\ and\ \bibinfo {author} {\bibfnamefont {I.~A.}\ \bibnamefont
  {Walmsley}},\ }\bibfield  {title} {\enquote {\bibinfo {title} {{Spectral
  phase interferometry for direct electric-field reconstruction of ultrashort
  optical pulses}},}\ }\href {\doibase 10.1364/OL.23.000792} {\bibfield
  {journal} {\bibinfo  {journal} {Optics Letters}\ }\textbf {\bibinfo {volume}
  {23}},\ \bibinfo {pages} {792} (\bibinfo {year} {1998})}\BibitemShut
  {NoStop}%
\bibitem [{\citenamefont {Miranda}\ \emph {et~al.}(2012)\citenamefont
  {Miranda}, \citenamefont {Arnold}, \citenamefont {Fordell}, \citenamefont
  {Silva}, \citenamefont {Alonso}, \citenamefont {Weigand}, \citenamefont
  {L'Huillier},\ and\ \citenamefont {Crespo}}]{Miranda:12}%
  \BibitemOpen
  \bibfield  {author} {\bibinfo {author} {\bibfnamefont {Miguel}\ \bibnamefont
  {Miranda}}, \bibinfo {author} {\bibfnamefont {Cord~L}\ \bibnamefont
  {Arnold}}, \bibinfo {author} {\bibfnamefont {Thomas}\ \bibnamefont
  {Fordell}}, \bibinfo {author} {\bibfnamefont {Francisco}\ \bibnamefont
  {Silva}}, \bibinfo {author} {\bibfnamefont {Benjam{\'{i}}n}\ \bibnamefont
  {Alonso}}, \bibinfo {author} {\bibfnamefont {Rosa}\ \bibnamefont {Weigand}},
  \bibinfo {author} {\bibfnamefont {Anne}\ \bibnamefont {L'Huillier}}, \ and\
  \bibinfo {author} {\bibfnamefont {Helder}\ \bibnamefont {Crespo}},\
  }\bibfield  {title} {\enquote {\bibinfo {title} {{Characterization of
  broadband few-cycle laser pulses with the d-scan technique}},}\ }\href
  {\doibase 10.1364/OE.20.018732} {\bibfield  {journal} {\bibinfo  {journal}
  {Opt. Express}\ }\textbf {\bibinfo {volume} {20}},\ \bibinfo {pages}
  {18732--18743} (\bibinfo {year} {2012})}\BibitemShut {NoStop}%
\bibitem [{\citenamefont {Boyd}(1992)}]{Boyd}%
  \BibitemOpen
  \bibfield  {author} {\bibinfo {author} {\bibfnamefont {Robert~W}\
  \bibnamefont {Boyd}},\ }\href@noop {} {\emph {\bibinfo {title} {{Nonlinear
  optics}}}}\ (\bibinfo  {publisher} {Academic Press},\ \bibinfo {address} {San
  Diego and London},\ \bibinfo {year} {1992})\BibitemShut {NoStop}%
\bibitem [{\citenamefont {Itatani}\ \emph {et~al.}(2002)\citenamefont
  {Itatani}, \citenamefont {Qu{\'{e}}r{\'{e}}}, \citenamefont {Yudin},
  \citenamefont {Ivanov}, \citenamefont {Krausz},\ and\ \citenamefont
  {Corkum}}]{ItataniPRL2002}%
  \BibitemOpen
  \bibfield  {author} {\bibinfo {author} {\bibfnamefont {J}~\bibnamefont
  {Itatani}}, \bibinfo {author} {\bibfnamefont {F}~\bibnamefont
  {Qu{\'{e}}r{\'{e}}}}, \bibinfo {author} {\bibfnamefont {G~L}\ \bibnamefont
  {Yudin}}, \bibinfo {author} {\bibfnamefont {M~Yu.}\ \bibnamefont {Ivanov}},
  \bibinfo {author} {\bibfnamefont {F}~\bibnamefont {Krausz}}, \ and\ \bibinfo
  {author} {\bibfnamefont {P~B}\ \bibnamefont {Corkum}},\ }\bibfield  {title}
  {\enquote {\bibinfo {title} {{Attosecond Streak Camera}},}\ }\href {\doibase
  10.1103/PhysRevLett.88.173903} {\bibfield  {journal} {\bibinfo  {journal}
  {Phys. Rev. Lett.}\ }\textbf {\bibinfo {volume} {88}},\ \bibinfo {pages}
  {173903} (\bibinfo {year} {2002})}\BibitemShut {NoStop}%
\bibitem [{\citenamefont {Goulielmakis}\ \emph {et~al.}(2004)\citenamefont
  {Goulielmakis}, \citenamefont {Uiberacker}, \citenamefont {Kienberger},
  \citenamefont {Baltuska}, \citenamefont {Yakovlev}, \citenamefont {Scrinzi},
  \citenamefont {Westerwalbesloh}, \citenamefont {Kleineberg}, \citenamefont
  {Heinzmann}, \citenamefont {Drescher},\ and\ \citenamefont
  {Krausz}}]{GoulielmakisScience2004}%
  \BibitemOpen
  \bibfield  {author} {\bibinfo {author} {\bibfnamefont {E}~\bibnamefont
  {Goulielmakis}}, \bibinfo {author} {\bibfnamefont {M}~\bibnamefont
  {Uiberacker}}, \bibinfo {author} {\bibfnamefont {R}~\bibnamefont
  {Kienberger}}, \bibinfo {author} {\bibfnamefont {A}~\bibnamefont {Baltuska}},
  \bibinfo {author} {\bibfnamefont {V}~\bibnamefont {Yakovlev}}, \bibinfo
  {author} {\bibfnamefont {A}~\bibnamefont {Scrinzi}}, \bibinfo {author}
  {\bibfnamefont {Th.}\ \bibnamefont {Westerwalbesloh}}, \bibinfo {author}
  {\bibfnamefont {U}~\bibnamefont {Kleineberg}}, \bibinfo {author}
  {\bibfnamefont {U}~\bibnamefont {Heinzmann}}, \bibinfo {author}
  {\bibfnamefont {M}~\bibnamefont {Drescher}}, \ and\ \bibinfo {author}
  {\bibfnamefont {F}~\bibnamefont {Krausz}},\ }\bibfield  {title} {\enquote
  {\bibinfo {title} {{Direct Measurement of Light Waves}},}\ }\href {\doibase
  10.1126/science.1100866} {\bibfield  {journal} {\bibinfo  {journal}
  {Science}\ }\textbf {\bibinfo {volume} {305}},\ \bibinfo {pages} {1267--1269}
  (\bibinfo {year} {2004})}\BibitemShut {NoStop}%
\bibitem [{\citenamefont {Kitzler}\ \emph {et~al.}(2002)\citenamefont
  {Kitzler}, \citenamefont {Milosevic}, \citenamefont {Scrinzi}, \citenamefont
  {Krausz},\ and\ \citenamefont {Brabec}}]{Kitzler2002}%
  \BibitemOpen
  \bibfield  {author} {\bibinfo {author} {\bibfnamefont {Markus}\ \bibnamefont
  {Kitzler}}, \bibinfo {author} {\bibfnamefont {Nenad}\ \bibnamefont
  {Milosevic}}, \bibinfo {author} {\bibfnamefont {Armin}\ \bibnamefont
  {Scrinzi}}, \bibinfo {author} {\bibfnamefont {Ferenc}\ \bibnamefont
  {Krausz}}, \ and\ \bibinfo {author} {\bibfnamefont {Thomas}\ \bibnamefont
  {Brabec}},\ }\bibfield  {title} {\enquote {\bibinfo {title} {{Quantum Theory
  of Attosecond XUV Pulse Measurement by Laser Dressed Photoionization}},}\
  }\href {\doibase 10.1103/PhysRevLett.88.173904} {\bibfield  {journal}
  {\bibinfo  {journal} {Physical Review Letters}\ }\textbf {\bibinfo {volume}
  {88}},\ \bibinfo {pages} {173904} (\bibinfo {year} {2002})}\BibitemShut
  {NoStop}%
\bibitem [{\citenamefont {Bethe}\ and\ \citenamefont
  {Salpeter}(1977)}]{bethe-salpeter}%
  \BibitemOpen
  \bibfield  {author} {\bibinfo {author} {\bibfnamefont {H~A}\ \bibnamefont
  {Bethe}}\ and\ \bibinfo {author} {\bibfnamefont {E~E}\ \bibnamefont
  {Salpeter}},\ }\enquote {\bibinfo {title} {{Quantum Mechanics of One- and
  Two-Electron Atoms}},}\ \ (\bibinfo  {publisher} {Plenum Publishing
  Corporation},\ \bibinfo {address} {New York},\ \bibinfo {year} {1977})\
  p.~\bibinfo {pages} {96}\BibitemShut {NoStop}%
\bibitem [{\citenamefont {Chini}\ \emph {et~al.}(2010)\citenamefont {Chini},
  \citenamefont {Gilbertson}, \citenamefont {Khan},\ and\ \citenamefont
  {Chang}}]{ChiniOE2010}%
  \BibitemOpen
  \bibfield  {author} {\bibinfo {author} {\bibfnamefont {Michael}\ \bibnamefont
  {Chini}}, \bibinfo {author} {\bibfnamefont {Steve}\ \bibnamefont
  {Gilbertson}}, \bibinfo {author} {\bibfnamefont {Sabih~D}\ \bibnamefont
  {Khan}}, \ and\ \bibinfo {author} {\bibfnamefont {Zenghu}\ \bibnamefont
  {Chang}},\ }\bibfield  {title} {\enquote {\bibinfo {title} {{Characterizing
  ultrabroadband attosecond lasers}},}\ }\href {\doibase 10.1364/OE.18.013006}
  {\bibfield  {journal} {\bibinfo  {journal} {Opt. Express}\ }\textbf {\bibinfo
  {volume} {18}},\ \bibinfo {pages} {13006--13016} (\bibinfo {year}
  {2010})}\BibitemShut {NoStop}%
\bibitem [{\citenamefont {Dahlstr{\"{o}}m}\ \emph
  {et~al.}(2012{\natexlab{a}})\citenamefont {Dahlstr{\"{o}}m}, \citenamefont
  {L'Huillier},\ and\ \citenamefont {Maquet}}]{DahlstromJPB2012}%
  \BibitemOpen
  \bibfield  {author} {\bibinfo {author} {\bibfnamefont {J~M}\ \bibnamefont
  {Dahlstr{\"{o}}m}}, \bibinfo {author} {\bibfnamefont {A}~\bibnamefont
  {L'Huillier}}, \ and\ \bibinfo {author} {\bibfnamefont {A}~\bibnamefont
  {Maquet}},\ }\bibfield  {title} {\enquote {\bibinfo {title} {{Introduction to
  attosecond delays in photoionization}},}\ }\href@noop {} {\bibfield
  {journal} {\bibinfo  {journal} {Journal of Physics B: Atomic, Molecular and
  Optical Physics}\ }\textbf {\bibinfo {volume} {45}},\ \bibinfo {pages}
  {183001} (\bibinfo {year} {2012}{\natexlab{a}})}\BibitemShut {NoStop}%
\bibitem [{\citenamefont {Feist}\ \emph {et~al.}(2014)\citenamefont {Feist},
  \citenamefont {Zatsarinny}, \citenamefont {Nagele}, \citenamefont {Pazourek},
  \citenamefont {Burgd{\"{o}}rfer}, \citenamefont {Guan}, \citenamefont
  {Bartschat},\ and\ \citenamefont {Schneider}}]{FeistPRA2014}%
  \BibitemOpen
  \bibfield  {author} {\bibinfo {author} {\bibfnamefont {Johannes}\
  \bibnamefont {Feist}}, \bibinfo {author} {\bibfnamefont {Oleg}\ \bibnamefont
  {Zatsarinny}}, \bibinfo {author} {\bibfnamefont {Stefan}\ \bibnamefont
  {Nagele}}, \bibinfo {author} {\bibfnamefont {Renate}\ \bibnamefont
  {Pazourek}}, \bibinfo {author} {\bibfnamefont {Joachim}\ \bibnamefont
  {Burgd{\"{o}}rfer}}, \bibinfo {author} {\bibfnamefont {Xiaoxu}\ \bibnamefont
  {Guan}}, \bibinfo {author} {\bibfnamefont {Klaus}\ \bibnamefont {Bartschat}},
  \ and\ \bibinfo {author} {\bibfnamefont {Barry~I}\ \bibnamefont
  {Schneider}},\ }\bibfield  {title} {\enquote {\bibinfo {title} {{Time delays
  for attosecond streaking in photoionization of neon}},}\ }\href {\doibase
  10.1103/PhysRevA.89.033417} {\bibfield  {journal} {\bibinfo  {journal} {Phys.
  Rev. A}\ }\textbf {\bibinfo {volume} {89}},\ \bibinfo {pages} {33417}
  (\bibinfo {year} {2014})}\BibitemShut {NoStop}%
\bibitem [{\citenamefont {Gu{\'{e}}not}\ \emph {et~al.}(2012)\citenamefont
  {Gu{\'{e}}not}, \citenamefont {Kl{\"{u}}nder}, \citenamefont {Arnold},
  \citenamefont {Kroon}, \citenamefont {Dahlstr{\"{o}}m}, \citenamefont
  {Miranda}, \citenamefont {Fordell}, \citenamefont {Gisselbrecht},
  \citenamefont {Johnsson}, \citenamefont {Mauritsson}, \citenamefont
  {Lindroth}, \citenamefont {Maquet}, \citenamefont {Ta{\"{i}}eb},
  \citenamefont {L'Huillier},\ and\ \citenamefont {Kheifets}}]{GuenotPRA2012}%
  \BibitemOpen
  \bibfield  {author} {\bibinfo {author} {\bibfnamefont {D}~\bibnamefont
  {Gu{\'{e}}not}}, \bibinfo {author} {\bibfnamefont {K}~\bibnamefont
  {Kl{\"{u}}nder}}, \bibinfo {author} {\bibfnamefont {C~L}\ \bibnamefont
  {Arnold}}, \bibinfo {author} {\bibfnamefont {D}~\bibnamefont {Kroon}},
  \bibinfo {author} {\bibfnamefont {J~M}\ \bibnamefont {Dahlstr{\"{o}}m}},
  \bibinfo {author} {\bibfnamefont {M}~\bibnamefont {Miranda}}, \bibinfo
  {author} {\bibfnamefont {T}~\bibnamefont {Fordell}}, \bibinfo {author}
  {\bibfnamefont {M}~\bibnamefont {Gisselbrecht}}, \bibinfo {author}
  {\bibfnamefont {P}~\bibnamefont {Johnsson}}, \bibinfo {author} {\bibfnamefont
  {J}~\bibnamefont {Mauritsson}}, \bibinfo {author} {\bibfnamefont
  {E}~\bibnamefont {Lindroth}}, \bibinfo {author} {\bibfnamefont
  {A}~\bibnamefont {Maquet}}, \bibinfo {author} {\bibfnamefont {R}~\bibnamefont
  {Ta{\"{i}}eb}}, \bibinfo {author} {\bibfnamefont {A}~\bibnamefont
  {L'Huillier}}, \ and\ \bibinfo {author} {\bibfnamefont {A~S}\ \bibnamefont
  {Kheifets}},\ }\bibfield  {title} {\enquote {\bibinfo {title}
  {{Photoemission-time-delay measurements and calculations close to the
  3{\$}s{\$}-ionization-cross-section minimum in Ar}},}\ }\href {\doibase
  10.1103/PhysRevA.85.053424} {\bibfield  {journal} {\bibinfo  {journal} {Phys.
  Rev. A}\ }\textbf {\bibinfo {volume} {85}},\ \bibinfo {pages} {53424}
  (\bibinfo {year} {2012})}\BibitemShut {NoStop}%
\bibitem [{\citenamefont {Palatchi}\ \emph {et~al.}(2014)\citenamefont
  {Palatchi}, \citenamefont {Dahlstr{\"{o}}m}, \citenamefont {Kheifets},
  \citenamefont {Ivanov}, \citenamefont {Canaday}, \citenamefont {Agostini},\
  and\ \citenamefont {DiMauro}}]{PalatchiJPB2014}%
  \BibitemOpen
  \bibfield  {author} {\bibinfo {author} {\bibfnamefont {Caryn}\ \bibnamefont
  {Palatchi}}, \bibinfo {author} {\bibfnamefont {J~M}\ \bibnamefont
  {Dahlstr{\"{o}}m}}, \bibinfo {author} {\bibfnamefont {A~S}\ \bibnamefont
  {Kheifets}}, \bibinfo {author} {\bibfnamefont {I~A}\ \bibnamefont {Ivanov}},
  \bibinfo {author} {\bibfnamefont {D~M}\ \bibnamefont {Canaday}}, \bibinfo
  {author} {\bibfnamefont {P}~\bibnamefont {Agostini}}, \ and\ \bibinfo
  {author} {\bibfnamefont {L~F}\ \bibnamefont {DiMauro}},\ }\bibfield  {title}
  {\enquote {\bibinfo {title} {{Atomic delay in helium, neon, argon and
  krypton}},}\ }\href {http://stacks.iop.org/0953-4075/47/i=24/a=245003}
  {\bibfield  {journal} {\bibinfo  {journal} {Journal of Physics B: Atomic,
  Molecular and Optical Physics}\ }\textbf {\bibinfo {volume} {47}},\ \bibinfo
  {pages} {245003} (\bibinfo {year} {2014})}\BibitemShut {NoStop}%
\bibitem [{\citenamefont {Gu{\'{e}}not}\ \emph {et~al.}(2014)\citenamefont
  {Gu{\'{e}}not}, \citenamefont {Kroon}, \citenamefont {Balogh}, \citenamefont
  {Larsen}, \citenamefont {Kotur}, \citenamefont {Miranda}, \citenamefont
  {Fordell}, \citenamefont {Johnsson}, \citenamefont {Mauritsson},
  \citenamefont {Gisselbrecht}, \citenamefont {Varj{\`{u}}}, \citenamefont
  {Arnold}, \citenamefont {Carette}, \citenamefont {Kheifets}, \citenamefont
  {Lindroth}, \citenamefont {L'Huillier},\ and\ \citenamefont
  {Dahlstr{\"{o}}m}}]{GuenotJPB2014}%
  \BibitemOpen
  \bibfield  {author} {\bibinfo {author} {\bibfnamefont {D}~\bibnamefont
  {Gu{\'{e}}not}}, \bibinfo {author} {\bibfnamefont {D}~\bibnamefont {Kroon}},
  \bibinfo {author} {\bibfnamefont {E}~\bibnamefont {Balogh}}, \bibinfo
  {author} {\bibfnamefont {E~W}\ \bibnamefont {Larsen}}, \bibinfo {author}
  {\bibfnamefont {M}~\bibnamefont {Kotur}}, \bibinfo {author} {\bibfnamefont
  {M}~\bibnamefont {Miranda}}, \bibinfo {author} {\bibfnamefont
  {T}~\bibnamefont {Fordell}}, \bibinfo {author} {\bibfnamefont
  {P}~\bibnamefont {Johnsson}}, \bibinfo {author} {\bibfnamefont
  {J}~\bibnamefont {Mauritsson}}, \bibinfo {author} {\bibfnamefont
  {M}~\bibnamefont {Gisselbrecht}}, \bibinfo {author} {\bibfnamefont
  {K}~\bibnamefont {Varj{\`{u}}}}, \bibinfo {author} {\bibfnamefont {C~L}\
  \bibnamefont {Arnold}}, \bibinfo {author} {\bibfnamefont {T}~\bibnamefont
  {Carette}}, \bibinfo {author} {\bibfnamefont {A~S}\ \bibnamefont {Kheifets}},
  \bibinfo {author} {\bibfnamefont {E}~\bibnamefont {Lindroth}}, \bibinfo
  {author} {\bibfnamefont {A}~\bibnamefont {L'Huillier}}, \ and\ \bibinfo
  {author} {\bibfnamefont {J~M}\ \bibnamefont {Dahlstr{\"{o}}m}},\ }\bibfield
  {title} {\enquote {\bibinfo {title} {{Measurements of relative photoemission
  time delays in noble gas atoms}},}\ }\href
  {http://stacks.iop.org/0953-4075/47/i=24/a=245602} {\bibfield  {journal}
  {\bibinfo  {journal} {Journal of Physics B: Atomic, Molecular and Optical
  Physics}\ }\textbf {\bibinfo {volume} {47}},\ \bibinfo {pages} {245602}
  (\bibinfo {year} {2014})}\BibitemShut {NoStop}%
\bibitem [{\citenamefont {Mansson}\ \emph {et~al.}(2014)\citenamefont
  {Mansson}, \citenamefont {Guenot}, \citenamefont {Arnold}, \citenamefont
  {Kroon}, \citenamefont {Kasper}, \citenamefont {Dahlstr{\"{o}}m},
  \citenamefont {Lindroth}, \citenamefont {Kheifets}, \citenamefont
  {L'Huillier}, \citenamefont {Sorensen},\ and\ \citenamefont
  {Gisselbrecht}}]{ManssonNP2014}%
  \BibitemOpen
  \bibfield  {author} {\bibinfo {author} {\bibfnamefont {Erik~P}\ \bibnamefont
  {Mansson}}, \bibinfo {author} {\bibfnamefont {Diego}\ \bibnamefont {Guenot}},
  \bibinfo {author} {\bibfnamefont {Cord~L}\ \bibnamefont {Arnold}}, \bibinfo
  {author} {\bibfnamefont {David}\ \bibnamefont {Kroon}}, \bibinfo {author}
  {\bibfnamefont {Susan}\ \bibnamefont {Kasper}}, \bibinfo {author}
  {\bibfnamefont {J~Marcus}\ \bibnamefont {Dahlstr{\"{o}}m}}, \bibinfo {author}
  {\bibfnamefont {Eva}\ \bibnamefont {Lindroth}}, \bibinfo {author}
  {\bibfnamefont {Anatoli~S}\ \bibnamefont {Kheifets}}, \bibinfo {author}
  {\bibfnamefont {Anne}\ \bibnamefont {L'Huillier}}, \bibinfo {author}
  {\bibfnamefont {Stacey~L}\ \bibnamefont {Sorensen}}, \ and\ \bibinfo {author}
  {\bibfnamefont {Mathieu}\ \bibnamefont {Gisselbrecht}},\ }\bibfield  {title}
  {\enquote {\bibinfo {title} {{Double ionization probed on the attosecond
  timescale}},}\ }\href@noop {} {\bibfield  {journal} {\bibinfo  {journal}
  {Nature Physics}\ }\textbf {\bibinfo {volume} {10}},\ \bibinfo {pages}
  {207--211} (\bibinfo {year} {2014})}\BibitemShut {NoStop}%
\bibitem [{\citenamefont {Wigner}(1955)}]{WignerPR1955}%
  \BibitemOpen
  \bibfield  {author} {\bibinfo {author} {\bibfnamefont {Eugene~P}\
  \bibnamefont {Wigner}},\ }\bibfield  {title} {\enquote {\bibinfo {title}
  {{Lower Limit for the Energy Derivative of the Scattering Phase Shift}},}\
  }\href {\doibase 10.1103/PhysRev.98.145} {\bibfield  {journal} {\bibinfo
  {journal} {Phys. Rev.}\ }\textbf {\bibinfo {volume} {98}},\ \bibinfo {pages}
  {145--147} (\bibinfo {year} {1955})}\BibitemShut {NoStop}%
\bibitem [{\citenamefont {Dahlstr{\"{o}}m}\ \emph {et~al.}(2013)\citenamefont
  {Dahlstr{\"{o}}m}, \citenamefont {Gu{\'{e}}not}, \citenamefont
  {Kl{\"{u}}nder}, \citenamefont {Gisselbrecht}, \citenamefont {Mauritsson},
  \citenamefont {L'Huillier}, \citenamefont {Maquet},\ and\ \citenamefont
  {Ta{\"{i}}eb}}]{DahlstromCP2013}%
  \BibitemOpen
  \bibfield  {author} {\bibinfo {author} {\bibfnamefont {J~M}\ \bibnamefont
  {Dahlstr{\"{o}}m}}, \bibinfo {author} {\bibfnamefont {D}~\bibnamefont
  {Gu{\'{e}}not}}, \bibinfo {author} {\bibfnamefont {K}~\bibnamefont
  {Kl{\"{u}}nder}}, \bibinfo {author} {\bibfnamefont {M}~\bibnamefont
  {Gisselbrecht}}, \bibinfo {author} {\bibfnamefont {J}~\bibnamefont
  {Mauritsson}}, \bibinfo {author} {\bibfnamefont {A}~\bibnamefont
  {L'Huillier}}, \bibinfo {author} {\bibfnamefont {A}~\bibnamefont {Maquet}}, \
  and\ \bibinfo {author} {\bibfnamefont {R}~\bibnamefont {Ta{\"{i}}eb}},\
  }\bibfield  {title} {\enquote {\bibinfo {title} {{Theory of attosecond delays
  in laser-assisted photoionization}},}\ }\href {\doibase
  http://dx.doi.org/10.1016/j.chemphys.2012.01.017} {\bibfield  {journal}
  {\bibinfo  {journal} {Chemical Physics}\ }\textbf {\bibinfo {volume} {414}},\
  \bibinfo {pages} {53--64} (\bibinfo {year} {2013})}\BibitemShut {NoStop}%
\bibitem [{\citenamefont {Nagele}\ \emph {et~al.}(2011)\citenamefont {Nagele},
  \citenamefont {Pazourek}, \citenamefont {Feist}, \citenamefont
  {Doblhoff-Dier}, \citenamefont {Lemell}, \citenamefont
  {T$\backslash$Hok{\'{e}}si},\ and\ \citenamefont
  {Burgd{\"{o}}rfer}}]{NagelePRA2011}%
  \BibitemOpen
  \bibfield  {author} {\bibinfo {author} {\bibfnamefont {S}~\bibnamefont
  {Nagele}}, \bibinfo {author} {\bibfnamefont {R}~\bibnamefont {Pazourek}},
  \bibinfo {author} {\bibfnamefont {J}~\bibnamefont {Feist}}, \bibinfo {author}
  {\bibfnamefont {K}~\bibnamefont {Doblhoff-Dier}}, \bibinfo {author}
  {\bibfnamefont {C}~\bibnamefont {Lemell}}, \bibinfo {author} {\bibfnamefont
  {K}~\bibnamefont {T$\backslash$Hok{\'{e}}si}}, \ and\ \bibinfo {author}
  {\bibfnamefont {J}~\bibnamefont {Burgd{\"{o}}rfer}},\ }\bibfield  {title}
  {\enquote {\bibinfo {title} {{Time-resolved photoemission by attosecond
  streaking: extraction of time information}},}\ }\href@noop {} {\bibfield
  {journal} {\bibinfo  {journal} {Journal of Physics B: Atomic, Molecular and
  Optical Physics}\ }\textbf {\bibinfo {volume} {44}},\ \bibinfo {pages}
  {81001} (\bibinfo {year} {2011})}\BibitemShut {NoStop}%
\bibitem [{\citenamefont {Pazourek}\ \emph {et~al.}(2012)\citenamefont
  {Pazourek}, \citenamefont {Feist}, \citenamefont {Nagele},\ and\
  \citenamefont {Burgd{\"{o}}rfer}}]{PazourekPRL2012}%
  \BibitemOpen
  \bibfield  {author} {\bibinfo {author} {\bibfnamefont {Renate}\ \bibnamefont
  {Pazourek}}, \bibinfo {author} {\bibfnamefont {Johannes}\ \bibnamefont
  {Feist}}, \bibinfo {author} {\bibfnamefont {Stefan}\ \bibnamefont {Nagele}},
  \ and\ \bibinfo {author} {\bibfnamefont {Joachim}\ \bibnamefont
  {Burgd{\"{o}}rfer}},\ }\bibfield  {title} {\enquote {\bibinfo {title}
  {{Attosecond Streaking of Correlated Two-Electron Transitions in Helium}},}\
  }\href {\doibase 10.1103/PhysRevLett.108.163001} {\bibfield  {journal}
  {\bibinfo  {journal} {Phys. Rev. Lett.}\ }\textbf {\bibinfo {volume} {108}},\
  \bibinfo {pages} {163001} (\bibinfo {year} {2012})}\BibitemShut {NoStop}%
\bibitem [{\citenamefont {Pazourek}\ \emph {et~al.}(2013)\citenamefont
  {Pazourek}, \citenamefont {Nagele},\ and\ \citenamefont
  {Burgd{\"{o}}rfer}}]{PazourekFD2013}%
  \BibitemOpen
  \bibfield  {author} {\bibinfo {author} {\bibfnamefont {Renate}\ \bibnamefont
  {Pazourek}}, \bibinfo {author} {\bibfnamefont {Stefan}\ \bibnamefont
  {Nagele}}, \ and\ \bibinfo {author} {\bibfnamefont {Joachim}\ \bibnamefont
  {Burgd{\"{o}}rfer}},\ }\bibfield  {title} {\enquote {\bibinfo {title}
  {{Time-resolved photoemission on the attosecond scale: opportunities and
  challenges}},}\ }\href {\doibase 10.1039/C3FD00004D} {\bibfield  {journal}
  {\bibinfo  {journal} {Faraday Discuss.}\ }\textbf {\bibinfo {volume} {163}},\
  \bibinfo {pages} {353--376} (\bibinfo {year} {2013})}\BibitemShut {NoStop}%
\bibitem [{\citenamefont {Lindroth}\ and\ \citenamefont
  {Dahlstr{\"{o}}m}(2017)}]{Lindroth2017}%
  \BibitemOpen
  \bibfield  {author} {\bibinfo {author} {\bibfnamefont {Eva}\ \bibnamefont
  {Lindroth}}\ and\ \bibinfo {author} {\bibfnamefont {Jan~Marcus}\ \bibnamefont
  {Dahlstr{\"{o}}m}},\ }\bibfield  {title} {\enquote {\bibinfo {title}
  {{Attosecond delays in laser-assisted photodetachment from closed-shell
  negative ions}},}\ }\href {\doibase 10.1103/PhysRevA.96.013420} {\bibfield
  {journal} {\bibinfo  {journal} {Physical Review A}\ }\textbf {\bibinfo
  {volume} {96}},\ \bibinfo {pages} {013420} (\bibinfo {year} {2017})},\
  \Eprint {http://arxiv.org/abs/1706.07217} {arXiv:1706.07217} \BibitemShut
  {NoStop}%
\bibitem [{\citenamefont {Pabst}\ and\ \citenamefont
  {Dahlstr{\"{o}}m}(2017)}]{PabstJPB2017}%
  \BibitemOpen
  \bibfield  {author} {\bibinfo {author} {\bibfnamefont {Stefan}\ \bibnamefont
  {Pabst}}\ and\ \bibinfo {author} {\bibfnamefont {Jan~Marcus}\ \bibnamefont
  {Dahlstr{\"{o}}m}},\ }\bibfield  {title} {\enquote {\bibinfo {title}
  {{Characterizing attosecond pulses in the soft x-ray regime}},}\ }\href
  {http://stacks.iop.org/0953-4075/50/i=10/a=104002} {\bibfield  {journal}
  {\bibinfo  {journal} {Journal of Physics B: Atomic, Molecular and Optical
  Physics}\ }\textbf {\bibinfo {volume} {50}},\ \bibinfo {pages} {104002}
  (\bibinfo {year} {2017})}\BibitemShut {NoStop}%
\bibitem [{\citenamefont {Dahlstr{\"{o}}m}\ \emph
  {et~al.}(2012{\natexlab{b}})\citenamefont {Dahlstr{\"{o}}m}, \citenamefont
  {Carette},\ and\ \citenamefont {Lindroth}}]{DahlstromPRA2012}%
  \BibitemOpen
  \bibfield  {author} {\bibinfo {author} {\bibfnamefont {J~M}\ \bibnamefont
  {Dahlstr{\"{o}}m}}, \bibinfo {author} {\bibfnamefont {T}~\bibnamefont
  {Carette}}, \ and\ \bibinfo {author} {\bibfnamefont {E}~\bibnamefont
  {Lindroth}},\ }\bibfield  {title} {\enquote {\bibinfo {title} {{Diagrammatic
  approach to attosecond delays in photoionization}},}\ }\href {\doibase
  10.1103/PhysRevA.86.061402} {\bibfield  {journal} {\bibinfo  {journal} {Phys.
  Rev. A}\ }\textbf {\bibinfo {volume} {86}},\ \bibinfo {pages} {61402}
  (\bibinfo {year} {2012}{\natexlab{b}})}\BibitemShut {NoStop}%
\bibitem [{\citenamefont {Moore}\ \emph {et~al.}(2011)\citenamefont {Moore},
  \citenamefont {Lysaght}, \citenamefont {Parker}, \citenamefont {van~der
  Hart},\ and\ \citenamefont {Taylor}}]{MoorePRA2011}%
  \BibitemOpen
  \bibfield  {author} {\bibinfo {author} {\bibfnamefont {L~R}\ \bibnamefont
  {Moore}}, \bibinfo {author} {\bibfnamefont {M~A}\ \bibnamefont {Lysaght}},
  \bibinfo {author} {\bibfnamefont {J~S}\ \bibnamefont {Parker}}, \bibinfo
  {author} {\bibfnamefont {H~W}\ \bibnamefont {van~der Hart}}, \ and\ \bibinfo
  {author} {\bibfnamefont {K~T}\ \bibnamefont {Taylor}},\ }\bibfield  {title}
  {\enquote {\bibinfo {title} {{Time delay between photoemission from the
  {\$}2p{\$} and {\$}2s{\$} subshells of neon}},}\ }\href {\doibase
  10.1103/PhysRevA.84.061404} {\bibfield  {journal} {\bibinfo  {journal} {Phys.
  Rev. A}\ }\textbf {\bibinfo {volume} {84}},\ \bibinfo {pages} {61404}
  (\bibinfo {year} {2011})}\BibitemShut {NoStop}%
\bibitem [{\citenamefont {Tzallas}\ \emph {et~al.}(2003)\citenamefont
  {Tzallas}, \citenamefont {Charalambidis}, \citenamefont {Papadogiannis},
  \citenamefont {Witte},\ and\ \citenamefont {Tsakiris}}]{Tzallas2003}%
  \BibitemOpen
  \bibfield  {author} {\bibinfo {author} {\bibfnamefont {P.}~\bibnamefont
  {Tzallas}}, \bibinfo {author} {\bibfnamefont {D.}~\bibnamefont
  {Charalambidis}}, \bibinfo {author} {\bibfnamefont {N.~A.}\ \bibnamefont
  {Papadogiannis}}, \bibinfo {author} {\bibfnamefont {K.}~\bibnamefont
  {Witte}}, \ and\ \bibinfo {author} {\bibfnamefont {G.~D.}\ \bibnamefont
  {Tsakiris}},\ }\bibfield  {title} {\enquote {\bibinfo {title} {{Direct
  observation of attosecond light bunching}},}\ }\href {\doibase
  10.1038/nature02091} {\bibfield  {journal} {\bibinfo  {journal} {Nature}\
  }\textbf {\bibinfo {volume} {426}},\ \bibinfo {pages} {267--271} (\bibinfo
  {year} {2003})}\BibitemShut {NoStop}%
\bibitem [{\citenamefont {Thomson}\ \emph {et~al.}(2013)\citenamefont
  {Thomson}, \citenamefont {Reid},\ and\ \citenamefont {Leburn}}]{Thomson2013}%
  \BibitemOpen
  \bibfield  {author} {\bibinfo {author} {\bibfnamefont {Robert~R.}\
  \bibnamefont {Thomson}}, \bibinfo {author} {\bibfnamefont {D.T}\ \bibnamefont
  {Reid}}, \ and\ \bibinfo {author} {\bibfnamefont {C.T}\ \bibnamefont
  {Leburn}},\ }\href {\doibase 10.1007/978-3-319-00017-6} {\emph {\bibinfo
  {title} {Scottish Graduate series}}},\ \bibinfo {number} {August}\ (\bibinfo
  {year} {2013})\ pp.\ \bibinfo {pages} {334--358}\BibitemShut {NoStop}%
\bibitem [{\citenamefont {Dudovich}\ \emph {et~al.}(2006)\citenamefont
  {Dudovich}, \citenamefont {Smirnova}, \citenamefont {Levesque}, \citenamefont
  {Mairesse}, \citenamefont {Ivanov}, \citenamefont {Villeneuve},\ and\
  \citenamefont {Corkum}}]{Dudovich2006}%
  \BibitemOpen
  \bibfield  {author} {\bibinfo {author} {\bibfnamefont {N.}~\bibnamefont
  {Dudovich}}, \bibinfo {author} {\bibfnamefont {O.}~\bibnamefont {Smirnova}},
  \bibinfo {author} {\bibfnamefont {J.}~\bibnamefont {Levesque}}, \bibinfo
  {author} {\bibfnamefont {Y.}~\bibnamefont {Mairesse}}, \bibinfo {author}
  {\bibfnamefont {M.~Yu.}\ \bibnamefont {Ivanov}}, \bibinfo {author}
  {\bibfnamefont {D.~M.}\ \bibnamefont {Villeneuve}}, \ and\ \bibinfo {author}
  {\bibfnamefont {P.~B.}\ \bibnamefont {Corkum}},\ }\bibfield  {title}
  {\enquote {\bibinfo {title} {{Measuring and controlling the birth of
  attosecond XUV pulses}},}\ }\href {\doibase 10.1038/nphys434} {\bibfield
  {journal} {\bibinfo  {journal} {Nature Physics}\ }\textbf {\bibinfo {volume}
  {2}},\ \bibinfo {pages} {781--786} (\bibinfo {year} {2006})}\BibitemShut
  {NoStop}%
\bibitem [{\citenamefont {Shafir}\ \emph {et~al.}(2012)\citenamefont {Shafir},
  \citenamefont {Soifer}, \citenamefont {Bruner}, \citenamefont {Dagan},
  \citenamefont {Mairesse}, \citenamefont {Patchkovskii}, \citenamefont
  {Ivanov}, \citenamefont {Smirnova},\ and\ \citenamefont
  {Dudovich}}]{ShafirNature2012}%
  \BibitemOpen
  \bibfield  {author} {\bibinfo {author} {\bibfnamefont {Dror}\ \bibnamefont
  {Shafir}}, \bibinfo {author} {\bibfnamefont {Hadas}\ \bibnamefont {Soifer}},
  \bibinfo {author} {\bibfnamefont {Barry~D}\ \bibnamefont {Bruner}}, \bibinfo
  {author} {\bibfnamefont {Michal}\ \bibnamefont {Dagan}}, \bibinfo {author}
  {\bibfnamefont {Yann}\ \bibnamefont {Mairesse}}, \bibinfo {author}
  {\bibfnamefont {Serguei}\ \bibnamefont {Patchkovskii}}, \bibinfo {author}
  {\bibfnamefont {Misha~Yu.}\ \bibnamefont {Ivanov}}, \bibinfo {author}
  {\bibfnamefont {Olga}\ \bibnamefont {Smirnova}}, \ and\ \bibinfo {author}
  {\bibfnamefont {Nirit}\ \bibnamefont {Dudovich}},\ }\bibfield  {title}
  {\enquote {\bibinfo {title} {{Resolving the time when an electron exits a
  tunnelling barrier}},}\ }\href {http://dx.doi.org/10.1038/nature11025}
  {\bibfield  {journal} {\bibinfo  {journal} {Nature}\ }\textbf {\bibinfo
  {volume} {485}},\ \bibinfo {pages} {343--346} (\bibinfo {year}
  {2012})}\BibitemShut {NoStop}%
\bibitem [{\citenamefont {Gaarde}\ \emph {et~al.}(2008)\citenamefont {Gaarde},
  \citenamefont {Tate},\ and\ \citenamefont {Schafer}}]{GaardeJPB2008}%
  \BibitemOpen
  \bibfield  {author} {\bibinfo {author} {\bibfnamefont {Mette~B}\ \bibnamefont
  {Gaarde}}, \bibinfo {author} {\bibfnamefont {Jennifer~L}\ \bibnamefont
  {Tate}}, \ and\ \bibinfo {author} {\bibfnamefont {Kenneth~J}\ \bibnamefont
  {Schafer}},\ }\bibfield  {title} {\enquote {\bibinfo {title} {{Macroscopic
  aspects of attosecond pulse generation}},}\ }\href
  {http://stacks.iop.org/0953-4075/41/i=13/a=132001} {\bibfield  {journal}
  {\bibinfo  {journal} {Journal of Physics B: Atomic, Molecular and Optical
  Physics}\ }\textbf {\bibinfo {volume} {41}},\ \bibinfo {pages} {132001}
  (\bibinfo {year} {2008})}\BibitemShut {NoStop}%
\bibitem [{\citenamefont {Friedrich}(2006)}]{Friedrich2006}%
  \BibitemOpen
  \bibfield  {author} {\bibinfo {author} {\bibfnamefont {H}~\bibnamefont
  {Friedrich}},\ }\href@noop {} {\emph {\bibinfo {title} {{Theoretical Atomic
  Physics}}}}\ (\bibinfo  {publisher} {Springer-Verlag},\ \bibinfo {address}
  {Berlin Heidelberg},\ \bibinfo {year} {2006})\BibitemShut {NoStop}%
\bibitem [{\citenamefont {Cooper}(1962)}]{cooper:1962}%
  \BibitemOpen
  \bibfield  {author} {\bibinfo {author} {\bibfnamefont {John~W}\ \bibnamefont
  {Cooper}},\ }\bibfield  {title} {\enquote {\bibinfo {title} {{Photoionization
  from Outer Atomic Subshells. A Model Study}},}\ }\href {\doibase
  10.1103/PhysRev.128.681} {\bibfield  {journal} {\bibinfo  {journal} {Phys.
  Rev.}\ }\textbf {\bibinfo {volume} {128}},\ \bibinfo {pages} {681--693}
  (\bibinfo {year} {1962})}\BibitemShut {NoStop}%
\bibitem [{\citenamefont {Fano}(1961)}]{fanoPR1961}%
  \BibitemOpen
  \bibfield  {author} {\bibinfo {author} {\bibfnamefont {U}~\bibnamefont
  {Fano}},\ }\bibfield  {title} {\enquote {\bibinfo {title} {{Effects of
  Configuration Interaction on Intensities and Phase Shifts}},}\ }\href@noop {}
  {\bibfield  {journal} {\bibinfo  {journal} {Phys. Rev.}\ }\textbf {\bibinfo
  {volume} {124}},\ \bibinfo {pages} {1866} (\bibinfo {year}
  {1961})}\BibitemShut {NoStop}%
\bibitem [{\citenamefont {Zhao}\ and\ \citenamefont {Lin}(2005)}]{ZhaoPRA2005}%
  \BibitemOpen
  \bibfield  {author} {\bibinfo {author} {\bibfnamefont {Z.~X.}\ \bibnamefont
  {Zhao}}\ and\ \bibinfo {author} {\bibfnamefont {C.~D.}\ \bibnamefont {Lin}},\
  }\bibfield  {title} {\enquote {\bibinfo {title} {Theory of laser-assisted
  autoionization by attosecond light pulses},}\ }\href {\doibase
  10.1103/PhysRevA.71.060702} {\bibfield  {journal} {\bibinfo  {journal} {Phys.
  Rev. A}\ }\textbf {\bibinfo {volume} {71}},\ \bibinfo {pages} {060702}
  (\bibinfo {year} {2005})}\BibitemShut {NoStop}%
\bibitem [{\citenamefont {Pabst}\ \emph {et~al.}(2016)\citenamefont {Pabst},
  \citenamefont {Lein},\ and\ \citenamefont {W{\"{o}}rner}}]{Pabst2016}%
  \BibitemOpen
  \bibfield  {author} {\bibinfo {author} {\bibfnamefont {Stefan}\ \bibnamefont
  {Pabst}}, \bibinfo {author} {\bibfnamefont {Manfred}\ \bibnamefont {Lein}}, \
  and\ \bibinfo {author} {\bibfnamefont {Hans~Jakob}\ \bibnamefont
  {W{\"{o}}rner}},\ }\bibfield  {title} {\enquote {\bibinfo {title} {{Preparing
  attosecond coherences by strong-field ionization}},}\ }\href {\doibase
  10.1103/PhysRevA.93.023412} {\bibfield  {journal} {\bibinfo  {journal}
  {Physical Review A - Atomic, Molecular, and Optical Physics}\ }\textbf
  {\bibinfo {volume} {93}},\ \bibinfo {pages} {023412} (\bibinfo {year}
  {2016})},\ \Eprint {http://arxiv.org/abs/1506.08929} {arXiv:1506.08929}
  \BibitemShut {NoStop}%
\bibitem [{\citenamefont {Dahlström}\ \emph {et~al.}(2017)\citenamefont
  {Dahlström}, \citenamefont {Pabst},\ and\ \citenamefont
  {Lindroth}}]{DahlstromJO2017}%
  \BibitemOpen
  \bibfield  {author} {\bibinfo {author} {\bibfnamefont {Jan~Marcus}\
  \bibnamefont {Dahlström}}, \bibinfo {author} {\bibfnamefont {Stefan}\
  \bibnamefont {Pabst}}, \ and\ \bibinfo {author} {\bibfnamefont {Eva}\
  \bibnamefont {Lindroth}},\ }\bibfield  {title} {\enquote {\bibinfo {title}
  {Attosecond transient absorption of a bound wave packet coupled to a smooth
  continuum},}\ }\href {http://stacks.iop.org/2040-8986/19/i=11/a=114004}
  {\bibfield  {journal} {\bibinfo  {journal} {Journal of Optics}\ }\textbf
  {\bibinfo {volume} {19}},\ \bibinfo {pages} {114004} (\bibinfo {year}
  {2017})}\BibitemShut {NoStop}%
\bibitem [{NIS()}]{NIST:database}%
  \BibitemOpen
  \href@noop {} {\enquote {\bibinfo {title} {{No Title}},}\ }\bibinfo
  {howpublished} {NIST {\{}{\{}{\}}{\{}$\backslash$textbackslash{\}}it Atomic
  Spectra Database{\{}{\}}{\}}. (WWW published at
  http://physics.nist.gov/cgi-bin/AtData/main{\{}{\_}{\}}asd.)}\BibitemShut
  {NoStop}%
\bibitem [{\citenamefont {Lindgren}\ and\ \citenamefont
  {Ros{\'{e}}n}(1974)}]{lindgren:74:casehfs}%
  \BibitemOpen
  \bibfield  {author} {\bibinfo {author} {\bibfnamefont {I}~\bibnamefont
  {Lindgren}}\ and\ \bibinfo {author} {\bibfnamefont {A}~\bibnamefont
  {Ros{\'{e}}n}},\ }\bibfield  {title} {\enquote {\bibinfo {title}
  {{Relativistic self-consistent-field calculations with applications to atomic
  hyperfine interaction. Part II}},}\ }\href@noop {} {\bibfield  {journal}
  {\bibinfo  {journal} {Case Studies In Atomic Physics}\ }\textbf {\bibinfo
  {volume} {4}},\ \bibinfo {pages} {150--196} (\bibinfo {year}
  {1974})}\BibitemShut {NoStop}%
\bibitem [{\citenamefont {Fano}(1985)}]{FanoPRA1985}%
  \BibitemOpen
  \bibfield  {author} {\bibinfo {author} {\bibfnamefont {U}~\bibnamefont
  {Fano}},\ }\bibfield  {title} {\enquote {\bibinfo {title} {Propensity rules:
  An analytical approach},}\ }\href@noop {} {\bibfield  {journal} {\bibinfo
  {journal} {Phys. Rev. A}\ }\textbf {\bibinfo {volume} {32}},\ \bibinfo
  {pages} {617} (\bibinfo {year} {1985})}\BibitemShut {NoStop}%
\end{thebibliography}

%merlin.mbs apsrev4-1.bst 2010-07-25 4.21a (PWD, AO, DPC) hacked
%Control: key (0)
%Control: author (0) dotless jnrlst
%Control: editor formatted (1) identically to author
%Control: production of article title (0) allowed
%Control: page (1) range
%Control: year (0) verbatim
%Control: production of eprint (0) enabled
%

\end{document}